\documentclass{article}

\usepackage{arxiv}

\usepackage[utf8]{inputenc} 
\usepackage[T1]{fontenc}    
\usepackage{hyperref}       
\usepackage{url}            
\usepackage{booktabs}       
\usepackage{amsfonts}       
\usepackage{nicefrac}       
\usepackage{microtype}      
\usepackage{lipsum}		
\usepackage{graphicx}
\usepackage[square,numbers]{natbib}
\usepackage{doi}
\usepackage{multirow}
\usepackage{subcaption}
\usepackage{comment}

\title{Quantum Key Distribution Networks -- Key Management: A Survey}

\author{ \href{https://orcid.org/0000-0002-7981-7739}{\includegraphics[scale=0.06]{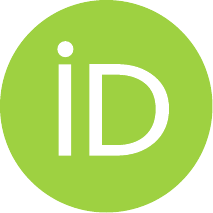}\hspace{1mm}Emir Dervisevic} \\
	Department of Telecommunications \\ Faculty of Electrical Engineering\\
	University of Sarajevo\\
	Sarajevo, Bosnia and Herzegovina \\
	\texttt{emir.dervisevic@etf.unsa.ba} \\
	\And
	{Amina Tankovic} \\
	Department of Telecommunications \\ Faculty of Electrical Engineering\\
	University of Sarajevo\\
	Sarajevo, Bosnia and Herzegovina \\
	\texttt{atankovic1@etf.unsa.ba} \\
        \AND
        {Ehsan Fazel} \\
        Cisco Quantum Lab \\
        Los Angeles, California, USA \\
        \texttt{salavifa@cisco.com} \\
        \And
        {Ramana Kompella} \\
        Cisco Quantum Lab \\
        Los Angeles, California, USA \\
        \texttt{rkompell@cisco.com} \\
        \And 
        {Peppino Fazio} \\
        DSMN, Ca' Foscari University of Venice \\
        Venice, Italy \\
        \texttt{peppino.fazio@unive.it} \\
        \And 
        {Miroslav Voznak} \\
        VSB – Technical University of Ostrava \\
        Ostrava, Czechia \\
        \texttt{miroslav.voznak@vsb.cz} \\
        \And 
        {Miralem Mehic} \\
        Department of Telecommunications \\ Faculty of Electrical Engineering\\
	University of Sarajevo\\
	Sarajevo, Bosnia and Herzegovina \\
        \texttt{miralem.mehic@ieee.org} \\
}

\renewcommand{\undertitle}{}
\renewcommand{\shorttitle}{Quantum Key Distribution Networks -- Key Management: A Survey}

\hypersetup{
pdftitle={Quantum Key Distribution Networks -- Key Management: A Survey},
pdfsubject={quantum key distribution, key management, networks},
pdfauthor={Emir Dervisevic, Amina Tankovic, Ehsan Fazel, Ramana Kompella, Peppino Fazio, Miroslav Voznak, Miralem Mehic},
pdfkeywords={quantum key distribution, network security, key management, survey},
}

\begin{document}
\maketitle

\begin{abstract}
    Secure communication makes the widespread use of telecommunication networks and services possible. With the constant progress of computing and mathematics, new cryptographic methods are being diligently developed. Quantum Key Distribution (QKD) is a promising technology that provides an Information-Theoretically Secure (ITS) solution to the secret-key agreement problem between two remote parties. QKD networks based on trusted repeaters are built to provide service to a larger number of parties at arbitrary distances. They function as an add-on technology to traditional networks, generating, managing, distributing, and supplying ITS cryptographic keys. Since key resources are limited, integrating QKD network services into critical infrastructures necessitates effective key management. As a result, this paper provides a comprehensive review of QKD network key management approaches. They are analyzed to facilitate the identification of potential strategies and accelerate the future development of QKD networks.
\end{abstract}


\section{Introduction}
\label{sec:introduction}
Secure means of communication are becoming increasingly important as data traffic in communication networks grows and more services emerge due to their integration~\cite{ullah20195g}. Sustaining widespread security mechanisms based on complex mathematical problems is proving challenging. Significant advances in computing and mathematics make it more challenging to ensure their security~\cite{grimes2019cryptography}. As a result, security experts have begun developing new cryptographic algorithms to address these challenges. The emergence of quantum computers is the most severe threat motivating these actions. Already-designed quantum algorithms pose a significant threat to public-key cryptosystems~\cite{shor1999polynomial}. It's just that there isn't a large-scale quantum computer to run them, at least not for practical applications.

Over the last two decades, tremendous efforts have been made to develop new cryptographic, quantum-secure mechanisms that will eventually replace the existing ones. They have resulted in two frameworks for secure communication in the post-quantum world: Post-Quantum Cryptography (PQC) and Quantum Cryptography. PQC concepts are based on a similar approach to classical algorithms: complex mathematical problems that cannot be solved in practical time by both classical and quantum computers~\cite{dam2023survey}. However, there is always the possibility that new quantum algorithms will be discovered in the near future, compromising their security. In contrast, quantum cryptography is based on the principles of quantum physics. Because the laws of quantum physics are unbreakable, the technology offers a long-term security solution. It is unaffected by advancements in computing or mathematics. However, it has considerable limitations.

Quantum Key Distribution (QKD)~\cite{bennett1984quantum} is the most mature example of quantum technologies. 
It has been in experimental testing for over two decades and has only recently been used in commercial applications. QKD is a method for agreeing on secret keys, a problem that cryptographers have long faced. The real advantage of QKD over traditional key-agreement protocols is that the established keys are Information-Theoretically Secure (ITS)~\cite{shor2000simple}. It is unique in many other aspects, including the method of implementation. QKD necessitates specialized hardware, whereas traditional mechanisms are typically implemented in software and use Internet services to negotiate a secret key.

\begin{figure} [h]
    \centering
    \begin{subfigure}{0.48\textwidth}
        \includegraphics[width=\textwidth]{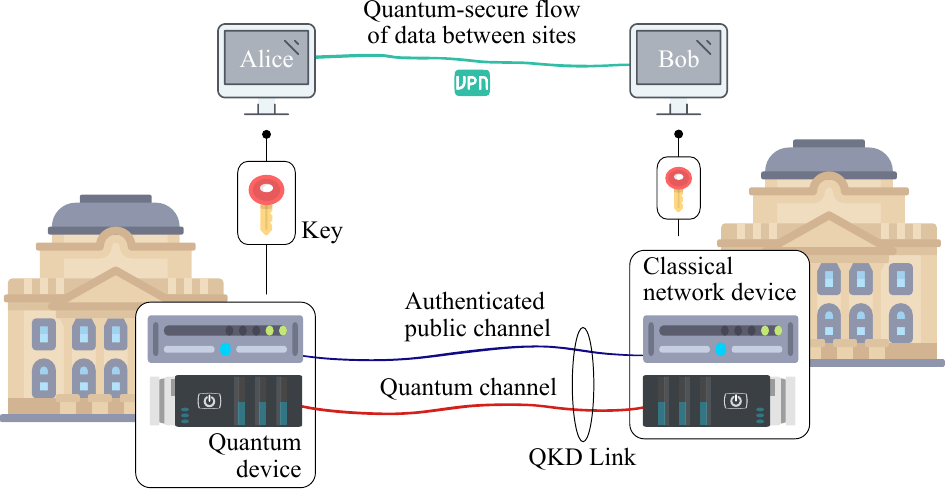}
        \caption{\label{fig:qkd-link}}
    \end{subfigure}
    \hfill
    \begin{subfigure}{0.48\textwidth}
        \includegraphics[width=\textwidth]{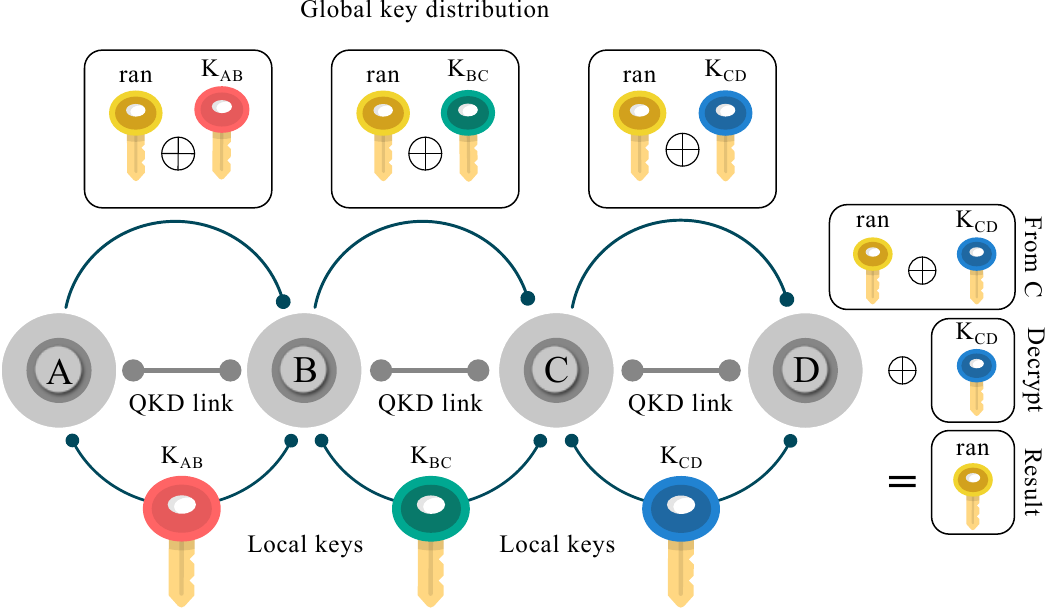}
        \caption{\label{fig:rhop}}
    \end{subfigure}

    \caption{a) Quantum key distribution between two remote sites. QKD-derived key material is used to establish quantum-secure flow of data between two applications~\cite{dervisevic2021overview}; b) Hop-by-hop global key distribution of a random key generated at a source node.}
    \label{fig:hop-relaying}
\end{figure}

QKD requires two channels: a quantum channel and an authenticated public channel, as shown in Figure~\ref{fig:qkd-link}. These two channels are commonly referred to as a logical QKD link.
Quantum transmission carried over the quantum channel cannot be passively monitored. When quantum carriers are monitored, they change state and, with high probability, reveal the presence of an eavesdropper. For determining if an eavesdropper is present and, if not, to correlate the data exchanged over the quantum channel, the authenticated public channel is required~\cite{bennett1992experimental, brassard1993secret, bennett1986reduce, bennett1988privacy}. The result is an ITS secret key, i.e., a true random sequence of bits known only to two legitimate parties. In the following discussion, these keys between two directly linked users are referred to as local keys. 

The traits that distinguish QKD from traditional approaches also add to the difficulties of large-scale deployment and application.
QKD, which requires a direct physical connection between two users, is primarily a point-to-point technology. Furthermore, the properties of quantum transmission prevent the use of classic amplifiers, limiting the range of the technology~\cite{alleaume2009topological}. 
As a result, QKD networks based on trusted repeaters are being built to achieve the goal of global QKD deployment, overcoming connectivity and distance limitations~\cite{elliott2002building}.
Key distribution or key relaying over a network of trusted repeaters is shown in Figure~\ref{fig:rhop}. The source node generates a random secret transmitted to the destination. The random secret, known as a global key, is One-Time-Pad (OTP) encrypted between each pair of trusted-repeater nodes using local keys. Assuming intermediate nodes are trusted and the random key is a genuinely random sequence of bits, the source and distant destination nodes establish an ITS global key. Instead of distributing a random key, the local key that the source shares with its first neighbor on route to the destination can be used. 

Although the majority of attention is still focused on the implementation of QKD itself, intending to achieve greater distances~\cite{yin2017satellite, liao2017satellite, wang2022twin, chen2022quantum, neumann2022continuous} and key rates~\cite{wang2022sub, grunenfelder2023fast, li2023high}, the level of attention devoted to the operation of QKD networks is gradually increasing~\cite{mehic2022quantum}. The establishment of testbeds worldwide~\cite{mehic2020quantum} has encouraged the development of key functionalities that must be addressed to achieve applicable technology. 

\subsection{Motivation}

Key management is one of the essential functionalities of QKD networks~\cite{itu3803} that is often neglected. 
However, the issue of effective key management must be addressed for the QKD network service to gain traction in modern telecommunications networks~\cite{mehic2023quantum}. A key manager is a device that manages keys. It performs various functions over keys, including key storage, key lifecycle management, key relaying, and key supply. Based on the comprehensive list of tasks, it is apparent that the key manager is a critical component of the QKD network infrastructure. It works with a finite quantity of cryptographic keys and provides them on request in accordance with the policy in place. As a result, service viability depends on effective key manager design. This article examines and compares existing designs in terms of supported functionalities. 

In this survey, we address the following questions:
\begin{itemize}
    \item How does the nature of a QKD process establish the need for key management?
    \item How does key management in the context of QKD networks differ from traditional approaches?
    \item What are the functional requirements for key management in a wide-scale QKD network?
    \item How have different testbeds and works approached the issue of key management? The discussion includes the approaches to key storage design, support for multiple applications etc.
    \item How do these different approaches compare to each other in terms of functionality?
    \item What are the key challenges that are still not addressed for functional and effective key management?
\end{itemize}

\renewcommand{\arraystretch}{1}
\begin{table}
\caption{Condensed compilation of terms encompassing prior survey research. Comparison to this survey.}
\label{tab:survey_comparison}
\centering
\resizebox{0.98\textwidth}{!}{%
\begin{tabular}{llccccc}
\toprule
\multicolumn{1}{c}{Reference} & \multicolumn{1}{c}{Year} & \begin{tabular}[c]{@{}c@{}}QKD\\ fundamentals\end{tabular} & \begin{tabular}[c]{@{}c@{}}QKD (point-to-point)\\ achievements\\ and implementations\end{tabular} & \begin{tabular}[c]{@{}c@{}}QKD network\\ fundamentals\end{tabular} & \begin{tabular}[c]{@{}c@{}}QKD\\ standardization\end{tabular} & \begin{tabular}[c]{@{}c@{}}QKD\\ key management\end{tabular} \\
\midrule
{\cite{alleaume2014using}}   & 2014  & \checkmark  & \checkmark  &   &   &            \\ \hline
{\cite{morris2014survey}}    & 2014  & \checkmark  & \checkmark  & \checkmark  &   &  \\ \hline
{\cite{diamanti2016practical}} & 2016  & \checkmark  & \checkmark   & \checkmark  & & \\ \hline
{\cite{geihs2019status}}       & 2019  & \checkmark  & \checkmark   & \checkmark   & \checkmark  &   \\ \hline
{\cite{xu2020secure}}          & 2020  & \checkmark  & \checkmark  &    &   &        \\ \hline
{\cite{cavaliere2020secure}}   & 2020  & \checkmark  & \checkmark    & \checkmark  & \checkmark   &            \\ \hline
{\cite{sharma2021quantum}}     & 2021  & \checkmark  & \checkmark    & \checkmark  &     &   \\ \hline
{\cite{amer2021introduction}}  & 2021  & \checkmark  & \checkmark    & \checkmark   &   &      \\ \hline
{\cite{tsai2021quantum}}       & 2021  & \checkmark  & \checkmark    & \checkmark   &   &      \\ \hline
{\cite{cao2022evolution}}      & 2022  & \checkmark  & \checkmark    & \checkmark   & \checkmark   &    \\ \hline
{\cite{mehic2023quantum}}       & 2023    & \checkmark   &   & \checkmark     &  \checkmark    &     \\ \hline
{Our survey}                     & 2024                      & \checkmark        &   & \checkmark   & \checkmark  & \checkmark                  \\
\bottomrule
\end{tabular}
}
\end{table}
\renewcommand{\arraystretch}{1}

\subsection{Comparison to Existing Surveys}
The surveys listed below and summarized in Table~\ref{tab:survey_comparison} cover the topic of quantum technologies:

\begin{itemize}
     \item Alleaume et al.~\cite{alleaume2014using} conducted an analysis and comparison of secret-key agreement techniques, with QKD being one of them, assessing its performance (recorded until 2014). Additionally, the study discusses approaches for constructing QKD networks and outlines two applications of QKD-derived key material in securing communication.
    \item Morris et al.~\cite{morris2014survey} examined different QKD protocols and network deployments, with a particular emphasis on the scale and performance of links.
    \item Diamanti et al.~\cite{diamanti2016practical} discussed QKD protocols and their experimental deployments, as well as approaches for constructing QKD networks.
    \item Geih et al.~\cite{geihs2019status} presented the achieved QKD performances, approaches for constructing QKD networks, and efforts towards standardization.
    \item Xu et al.~\cite{xu2020secure} examined the security of practical implementations of QKD using realistic flawed devices.
    \item Cavaliere et al.~\cite{cavaliere2020secure} presented the achieved link performances and challenges of QKD running within the same fiber as classical channels. Furthermore, the study discusses the physical characteristics of optical components and it briefly examines represented network deployments and ongoing standardization efforts.
    \item Sharma et al.~\cite{sharma2021quantum} analyzed QKD protocols and methods of integrating the technology into optical networks, addressing issues such as wavelength and time slot assignment for the quantum channel. The study further discusses networking aspects and summarizes existing real-world integrations of QKD into optical networks.
    \item Amer et al.~\cite{amer2021introduction} discussed various QKD protocols and their implementations, as well as established manufacturers in the field of QKD. Furthermore, the study briefly explains QKD network architecture and represented testbeds worldwide.
    \item Tsai et al.~\cite{tsai2021quantum} explore networking aspects and summarize key results achieved with represented testbeds. The primary focus is on network structure, achieved key rates and distances, and routing.
    \item Cao et al.~\cite{cao2022evolution} provided a comprehensive overview of state-of-the-art QKD protocols, performance, and practices for integrating quantum channels with classical channels within the same optical infrastructure. The study also describes QKD network architecture and essential building blocks. However, the key management layer is only briefly described in terms of basic requirements and lacks detailed analysis. Additionally, the survey covers ongoing progress in QKD standardization. 
    \item Mehic et al.~\cite{mehic2023quantum} conducted a survey on the integration of QKD with 5G networks. The study covers the fundamentals of QKD protocols and networking, discusses the integration of QKD in optical networks, and briefly touches on standardization efforts.
\end{itemize}

Based on the selected surveys, the narrowed list of research objectives discussed includes:
\begin{itemize}
    \item Analyzing QKD protocols and their implementations.
    \item Examining the state-of-the-art advancements in QKD technology and network deployments.
    \item Exploring real-world applications of QKD networks.
    \item Summarizing standardization efforts related to QKD protocols and networks.
\end{itemize}

A thorough examination of the literature, however, reveals the lack of a comprehensive system-view engineering perspective on QKD key management. To our knowledge, no references in the available literature thoroughly investigate the approaches to addressing key management issues. Consequently, this survey offers a chronological overview of existing approaches to the realization of key managers, unveiling the evolution of their functionality. This evolution begins with the need for simple key storage and progresses to sophisticated mechanisms aimed at improving efficiency and providing reliable service to a larger number of users. The identification of basic approaches and contributions from existing solutions led to the identification of existing gaps. Existing approaches were compared and analyzed to identify suitable approaches for developing an efficient key management system. 

This survey provides interested readers with a high-level system engineering viewpoint on the QKD network and its organization. Researchers, practitioners of quantum technology, and PhD students in the field of applied networking security will benefit from this survey and synthesis of perspectives on the confluence of modern technologies.

\subsection{Contribution}

The major contributions of this survey are outlined as follows:
\begin{enumerate}
    \item Offering a comprehensive insight of the operation and application of QKD networks through the perspective of key management.
    \item Providing a detailed overview of the functionality of key managers and elucidating the reasons for their existence.
    \item An overview of existing approaches to implementing key manager functionality.
    \item Unraveling the networking details of existing testbeds that are usually missing from the review literature.
    \item Discussion and analysis of existing solutions to identify suitable approaches for key management in QKD networks.
    \item Outlining the existing gaps in the research to provide clear guidelines for future work in the field of QKD key management.
\end{enumerate}

\subsection{Paper Organization}

The article is organized as follows. Section~\ref{sec:key-mngnt} describes key management issues and highlights the distinctions between QKD and classic key management. Section~\ref{sec:key-manager-solutions} provides a comprehensive review of existing key management solutions in chronological order of appearance. Section~\ref{sec:discussion} provides a comparative analysis of existing solutions, taking into account several key requirements and functionalities of key managers. This discussion has led to the identification of key challenges, which are outlined as guidelines for future directions in Section~\ref{sec:challenges}. Section~\ref{sec:conclusion} concludes our study.  

\section{Key Management} 
\label{sec:key-mngnt}
\begin{figure} [h]
    \centering
    \begin{subfigure}{0.35\textwidth}
        \includegraphics[width=\textwidth]{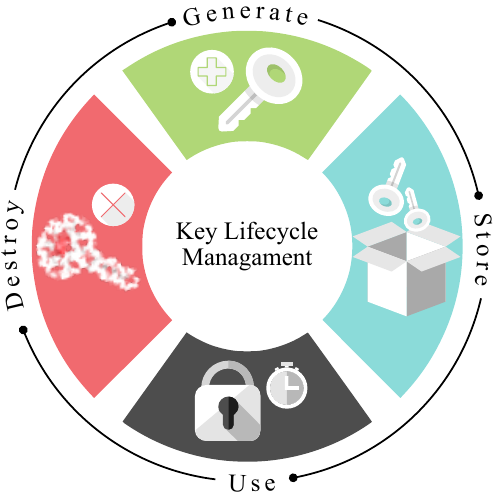}
        \caption{\label{fig:lifecycle}}
    \end{subfigure}
    \hfill
    \begin{subfigure}{0.55\textwidth}
        \includegraphics[width=\textwidth]{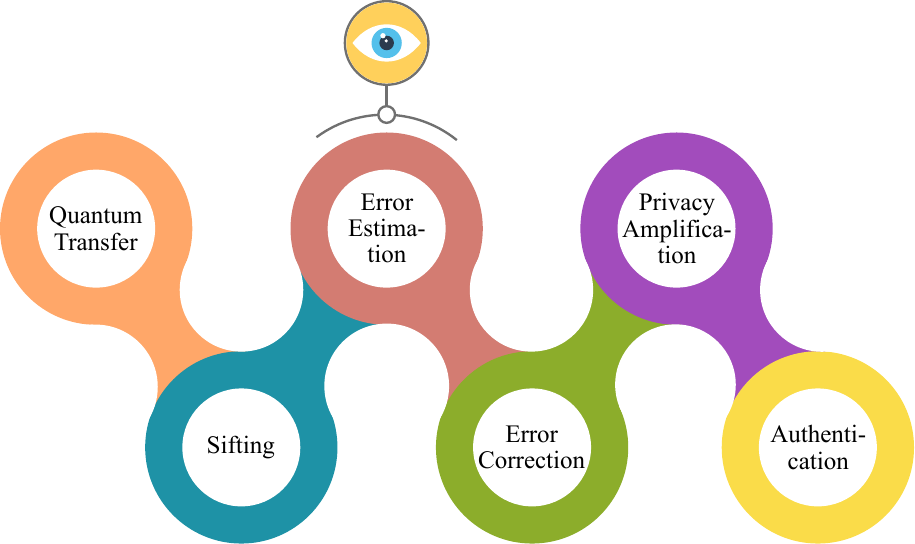}
        \caption{\label{fig:process}}
    \end{subfigure}

    \caption{a) Most essential cryptographic key life cycle management; b) QKD process is performed in several sequential steps. The process begins with the quantum transmission of a random sequence of key bits. 
    The correctness of information received through quantum transmission is highly dependent on measurement. Information obtained from incompatible measurements is discarded during a sifting procedure. The next step involves estimating the error rate and using its value to discover the eavesdropper. If the error rate is less than the threshold, the process proceeds to the error correction step. At the very end, a privacy amplification step is performed. The entire process must be authenticated.}
    \label{fig:lifecycle-process}
\end{figure}

Key management is a widely recognized concept in nearly all systems that support secure communication services. One framework we can use as an example is IPsec (Internet Protocol security), which manages the cryptographic keys needed to establish Virtual Private Networks (VPNs)~\cite{atkinson1995security}. To establish a secure VPN tunnel, the two peers must negotiate a key using the Diffie-Hellman key exchange protocol~\cite{kaufman2010internet}. The established key must be kept safe for the duration of its lifetime. The lifetime is expressed in data units (bytes) or elapsed time (seconds). It specifies how long a key can be used before it expires. Using an expired key is not recommended because it reduces encryption security and puts previously transmitted data at risk. Expired keys must be properly destroyed to prevent disclosure to third-party attackers who may use the "store now, decrypt later" strategy to decrypt harvested data using leaked keys. Once the keys have been destroyed, new ones should be established to continue the encryption of the data flow. Figure~\ref{fig:lifecycle} illustrates this essential key life cycle management and is recognized in IPsec and all security frameworks.

In contrast, the key management issues in QKD networks are highly distinct due to the intrinsic uniqueness of the QKD process. If QKD worked similarly to traditional key establishment techniques, it would replace the Diffie-Hellman key exchange from the earlier example. 
However the QKD process consists of several steps, as illustrated in Figure~\ref{fig:process}, and it could take several minutes until the process outputs ITS keys.
This is primarily why key management in QKD networks is a critical enabler of its services: generate larger amounts of cryptographic keys ahead of time and then supply them in a timely manner on demand to traditional security frameworks such as IPsec.

\begin{figure}[h]
    \centering
    \includegraphics[width= 0.98 \textwidth]{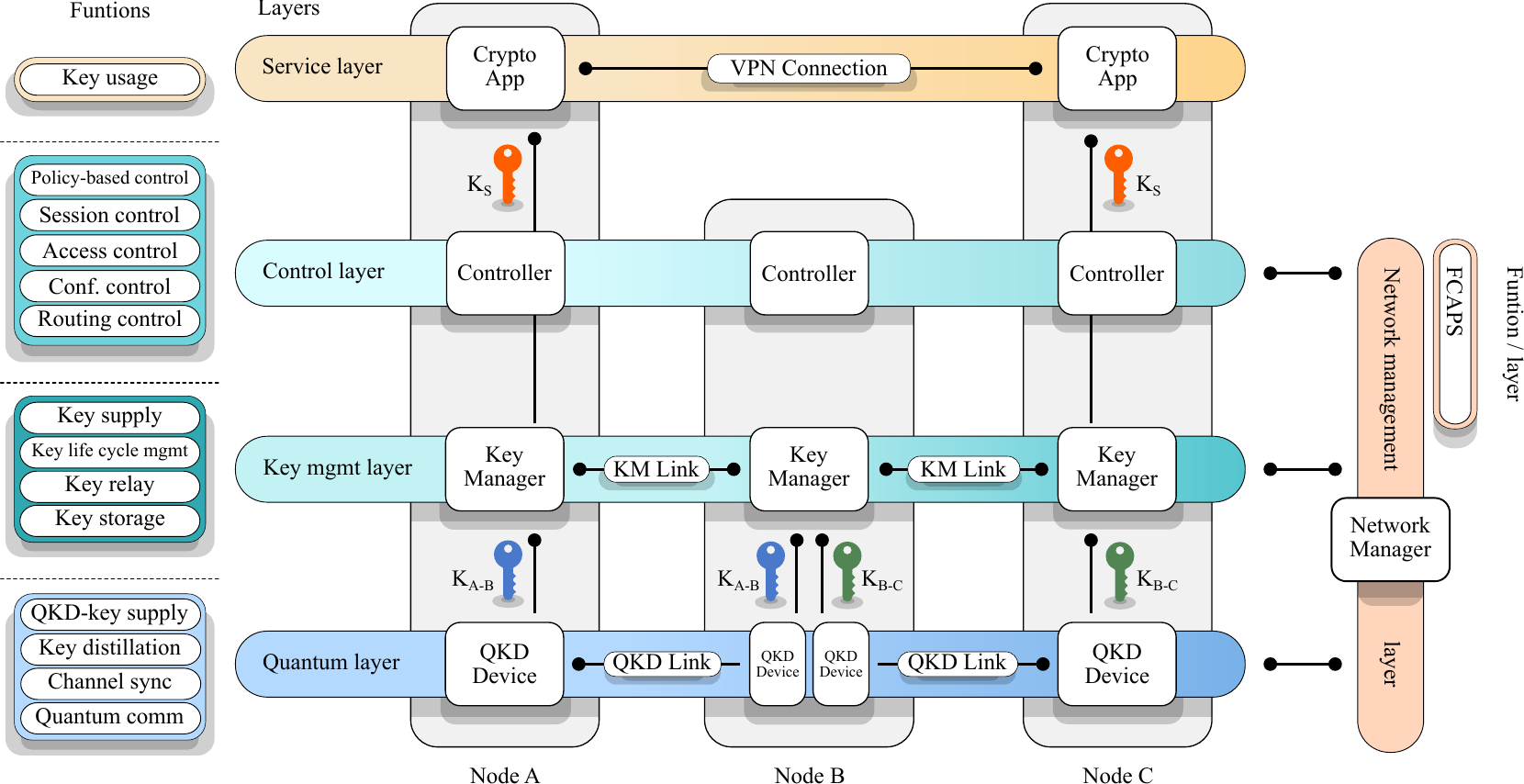}
    \caption{The layered architecture of QKD network.\label{fig:layered-arch}}
\end{figure}

Key management sits in the middle of the layers of the QKD network architecture as illustrated in Figure~\ref{fig:layered-arch}~\cite{itu3800}.
All the layers presented are fundamentally responsible for or influence key management in some manner.~\cite{itu3801}. 
However, while discussing key management in the context of QKD networks, one must surely refer to the key management layer. Only the key management layer can access cryptographic keys; other layers can only affect the key management strategy or gather key metadata, but not the key value.

At the key management layer, there is a functional element known as Key Manager (KM). It is also known as a Key Management System (KMS) in the literature. The ITU-T recommendation~\cite{itu3803} divides the functional requirements of KM into two separate agents: Key Management Agent (KMA) and Key Supply Agent (KSA). Due to their distinct key management tasks, this separation is defined for practical reasons. They can be installed on different machines within the same secure environment. The KMA includes the following functions: secure key storage, global key distribution, and key life cycle management, whereas the KSA includes a key supply function.

\subsection{Secure Key Storage}
\label{sec:storage}

In contrast to traditional cryptographic methods for secret key exchange, which are performed relatively on demand, QKD requires a significant time window and is thus performed in advance regardless of demand on cryptographic keys. This is a well-accepted practice of decoupling key generation and key consumption processes. This is achieved by introducing secure key storage. 
The key storage regularly receives fresh cryptographic keys. This event is managed by a quantum layer, which should ensure reliable and uninterrupted operation that delivers keys at a steady pace. 
The lengths of cryptographic keys produced at the quantum layer by the same or different QKD devices will vary. It is recommended that KM reformat keys to a specific unit length~\cite{itu3803}. This should be done before storage and requires interaction between involved KMs, as discussed in Section~\ref{sec:synchronization}.

Keys are consumed from the key storage at variable rates. The consumption process is driven by the number of cryptographic applications and their encryption preferences.
The way applications access keys is defined by the ETSI key delivery interfaces, ETSI QKD 014~\cite{etsi014} and ETSI QKD 004~\cite{etsi004}, which are described in Section~\ref{sec:supply}. 
To meet application requirements, KM may modify key entries (key splitting or merging) and assign unique key identifiers on supply. As a result, interaction between the KMs involved is required.

Decoupling key generation and key consumption processes addresses not only the large gap between two consecutive key generation events but also the QKD's limited key generation rates. Secure key storage allows the accumulation of larger amounts of cryptographic keys when there are low demands for consumption processes. A burst of high consumption demands that exceed the key generation rate can then be accommodated using existing key supplies.  

\subsection{Key synchronization}
\label{sec:synchronization}

QKD network as a service generates, manages, and supplies symmetric cryptographic keys. As a result, it is essential to maintain synchronization, or consistency, among the contents of key storages. Otherwise, the service is not operational. 
Even if the perfect correlation between two symmetric keys is proven at the quantum layer, the key management layer must verify that both KM peers receive these keys without errors. This is accomplished by exchanging message authentication codes, hash values calculated on key bits and identifiers. However, the security details of this verification are said to be outside of ITU-T recommendations~\cite{itu3803}.  

As it is anticipated that keys produced by different vendors will vary in size, it is recommended that the KM resizes keys to a specific unit of length before storing them. 
To resize keys, KMs must agree on a unit of length (which is typically set in advance) and unique identifiers for newly produced keys.\footnote{~The quantum layer generates large blocks (in the order of Mbits) of truly random bits. As a result, the resize operation will generally split this large key block into smaller, easier-to-manage blocks.} 
Given the distinct purpose of QKD networks, which involves the production, management, and supply of ITS keys, the security measures applied for synchronization purposes between remote Key Managers should be implemented with equally high levels of security. 
Synchronization messages can be frequent, leading to high internal consumption of keys at the key management layer.

The high frequency of synchronization messages is also a result of the creation and delivery of keys to the service layer. Cryptographic applications request keys with varying requirements, as discussed in Section~\ref{sec:supply}. 
Typically, one or more keys from the key storage are used to create the supply key. To accomplish this, key splitting or merging is used. The supply key is then given a unique identifier and supplied in response to the request. This modification of key entries and creation of supply key must be propagated to the peer KM. Similarly, the peer KM creates the supply key and waits until the respective cryptographic application pair requests it. 

It was previously stated that keys ought to be stored in a specific unit of length, but it didn't specify what that length should be. It can be predetermined, or, as recommended by ITU-T~\cite{itusupp70}, a machine learning mechanism can be used to learn about the application requirements and dynamically choose this specific length. The latter simplifies key supply because key transformation of available keys in storage is no longer required. Nevertheless, assigning a distinct key identifier and synchronizing it at the key management layer is still necessary. This signalization can help detect malicious requests~\cite{dervisevic2022simulations,mehic2022tackling} quickly.

\subsection{Global key distribution}
\label{sec:relay}

Given the uniqueness of the QKD process, which result in point-to-point connectivity with a limited distance, global key distribution allows peers to establish keys even if a QKD link does not directly connect them. One common method of global key distribution—hop-by-hop key distribution was covered in the introductory Section~\ref{sec:introduction}. Distributing random or local key in the hop-by-hop fashion pose a security risk because global keys are directly accessible to nodes along the distribution path.
As a result, apart from these intuitive approaches, the ITU-T~\cite{itu3803} recommends two other global key distribution options: distribution with XORs uniformly processed at the destination node and distribution with XORs collected at a single centralized node. A recent study proposes a modified centralized approach that relaxes the requirements of secure key storage at the intermediate nodes~\cite{dong2019wide}. Acknowledging the inherent single point of failure in centralized approaches, a distributed scheme has been introduced~\cite{vyas2024relaxing}. This distributed scheme maintains relaxed trust requirements while mitigating the risks associated with a centralized approach.

The global key distribution process, like the QKD process, takes time. For example, hop-by-hop key distribution involves multiple encryptions and decryptions that are both computationally and time-consuming. It also includes propagation and processing time. For a path with more nodes, the global key distribution can result in significant supply delays. 
As a result, the global key distribution process and key consumption process are separated. This is analogous to separating key generation and key consumption processes discussed in Section~\ref{sec:storage}. A node may distribute enough global keys (for a given destination) in advance to meet demands. 

\subsection{Key supply}
\label{sec:supply}

The key-supply interface describes access to the QKD network services by defining communication between cryptographic applications and KMs. ETSI has recently standardized two such interfaces: ETSI GS QKD 014~\cite{etsi014} and ETSI GS QKD 004~\cite{etsi004}.
Through the ETSI specification, cryptographic applications are also called Secure Application Entities (SAEs), while KMs are referred to as Key Management Entities (KMEs).

\subsubsection{ETSI GS QKD 014}
\label{sec:etsi_014}

\begin{figure} [h]
    \centering
    \begin{subfigure}{0.45\textwidth}
        \includegraphics[width=\textwidth]{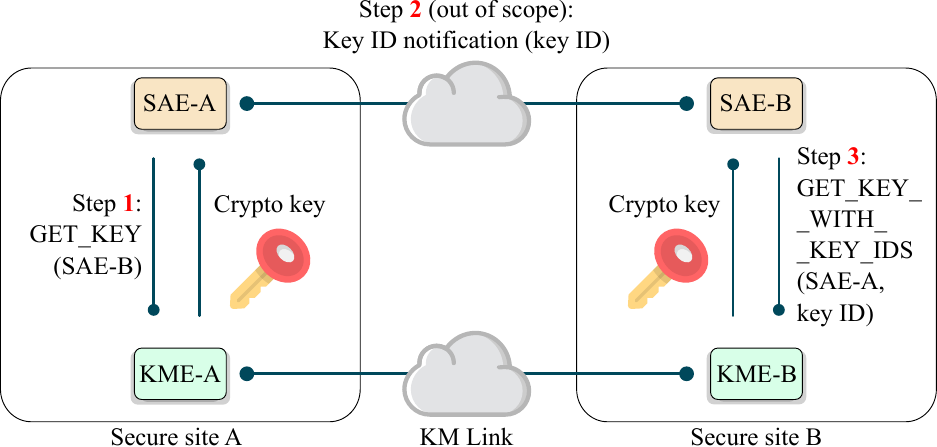}
        \caption{\label{fig:etsi-014}}
    \end{subfigure}
    \hfill
    \begin{subfigure}{0.45\textwidth}
        \includegraphics[width=\textwidth]{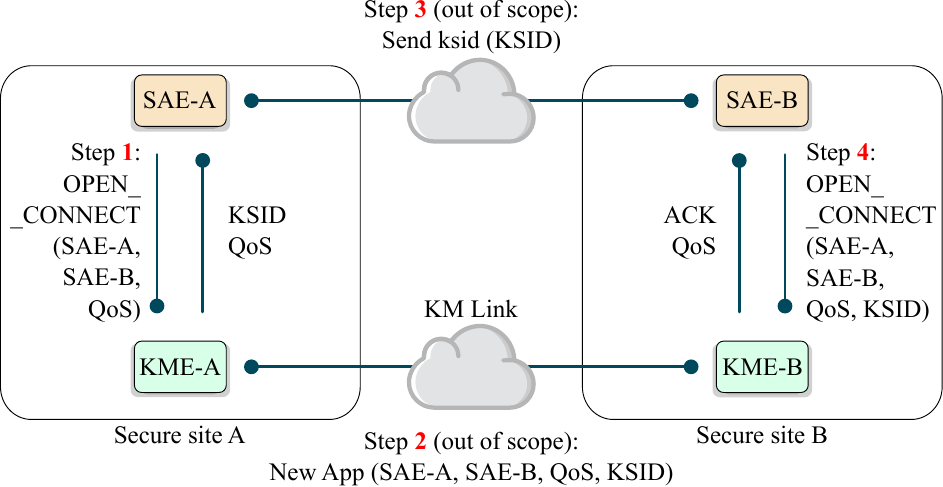}
        \caption{\label{fig:etsi-004}}
    \end{subfigure}

    \caption{a) Use-case of the ETSI QKD 014 key supply interface; b) Use-case of the ETSI QKD 004 key supply interface. Only key stream session establishment is shown.}
    \label{fig:etsi-sub}
\end{figure}

The ETSI QKD 014 specification defines an interface based on the HTTPS protocol and the JSON-encoded data format of posted parameters and responses. Because of its REST-based nature and simplified processing logic required at KMEs, this interface is well accepted among vendors who supply quantum equipment.
Three fundamental methods are used to depict the communication between SAEs and KMEs: GET\_STATUS, GET\_KEY, and GET\_KEY\_WITH\_KEY\_IDS. 
Figure~\ref{fig:etsi-014} depicts the use-case of the ETSI QKD 014 key-supply interface.
The GET\_STATUS method allows one SAE, let us call it SAE-A, to collect status information on QKD connection (which may be direct or virtual -- established through trusted repeaters) to a specific destination SAE-B. 
SAE-A obtains one or more keys from KME through GET\_KEY method. The request expresses the SAE's requirements by including the number and size of keys requested, additional destination SAEs, and other specific parameters. Provisioned keys are assigned unique key IDs that are synchronized between KMEs. After receiving key IDs through arbitrary means,\footnote{~The ETSI documentation does not describe how connecting SAEs communicate key IDs.} SAE-B uses the GET\_KEY\_WITH\_KEY\_IDS method to obtain the same set of keys.

\subsubsection{ETSI GS QKD 004}
\label{sec:etsi_004}

The ETSI GS QKD 004 introduces the concept of sessions with QoS capabilities for SAEs without requiring a specific protocol. 
Figure~\ref{fig:etsi-004} depicts the use-case of the ETSI QKD 004 key-supply interface.
Three primitive functions are defined: OPEN\_CONNECT, GET\_KEY, and CLOSE. The OPEN\_CONNECT function establishes a key stream session with the expected level of service for SAE. Establishing a key stream session requires KM to rendezvous with the designated destination KM. In the case of virtual QKD connections, the responsibility of KM is also to discover the destination KM and QoS available in the path of multiple KMs. This, however, is outside the scope of ETSI QKD 004. Once the key stream session is established or permitted, the calling SAE, for example, SAE-A, is granted a Key Stream IDentifier (KSID) used in subsequent key requests. To maintain the promised level of service, KM is responsible for managing and reserving keys for active key stream sessions. The QoS parameters that can be agreed upon are key size, maximum and minimum (requesting) key rate in bits per second (bps), jitter of key delivery, and priority level. The GET\_KEY function returns the required amount of key material requested for specific KSID while the CLOSE function allows SAEs to end and terminate key stream sessions. 

Another interesting notion discussed in ETSI QKD 004 is the organization of the key management layer. Each QKD module has its own key management unit, and a higher-level Key Server communicates with multiple QKD modules within the QKD node. It is emphasized that the ETSI QKD 004 interface is suitable for communication between key managers at various hierarchical levels. With multiple vendors pushing their proprietary KMs, this organization is probably the most likely to be implemented. Interoperability will also be attained through a new interface, ETSI QKD 020~\cite{etsi020}, which is currently in draft and describes horizontal communication between two KMs within the same trusted node, allowing one KM to pass the key to the other to achieve relay through this node.

\subsubsection{Cisco SKIP protocol}
\label{sec:skip}

\begin{figure} [h]
    \centering
    \begin{subfigure}{0.45\textwidth}
        \includegraphics[width=\textwidth]{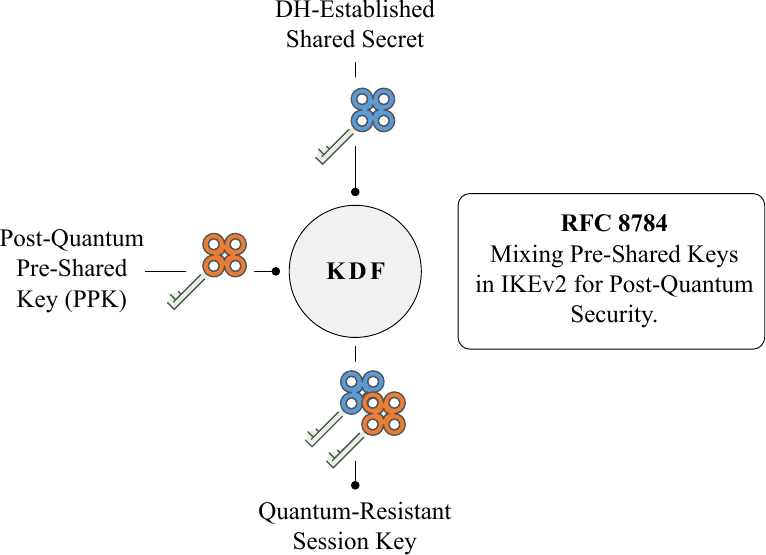}
        \caption{\label{fig:cisco-ikev2}}
    \end{subfigure}
    \hfill
    \begin{subfigure}{0.45\textwidth}
        \includegraphics[width=\textwidth]{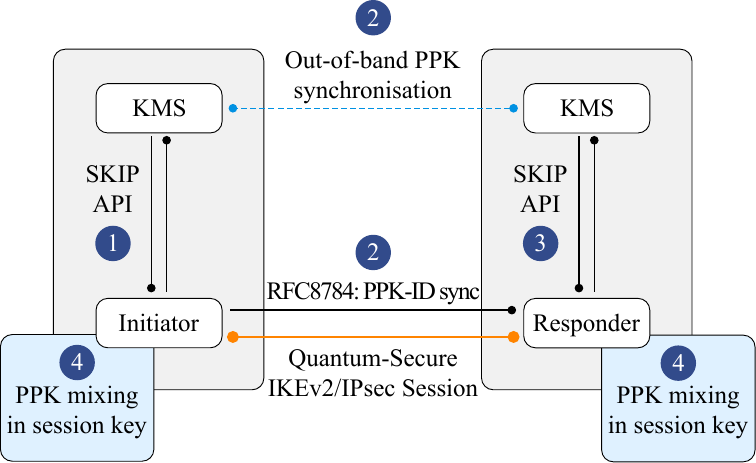}
        \caption{\label{fig:ppk}}
    \end{subfigure}

    \caption{a) The IKEv2 key derivation function - a mixture of the traditional key and the PPK; b) Quantum-Safe IKEv2 and IPsec Session Keys with a dynamic PPK as proposed by the Cisco}
    \label{fig:cisco-sub}
\end{figure}

Apart from the key-supply interfaces defined by the ETSI that were previously discussed, it is noteworthy to mention that certain commercial solutions are also available, like the Cisco Secure Key Integration Protocol (SKIP).
The Cisco IOS-XE relies on the enhanced IKEv2 protocol (RFC 8784~\cite{rfc8784}), which uses a mixture of traditional Elliptic Curve Diffie-Hellman (ECDH) cryptographic keys and Postquantum Preshared Keys (PPK) in the key derivation function. Figure~\ref{fig:cisco-ikev2} depicts the process of creating session keys. This feature enables quantum-safe encryption using PPKs and can be applied to all IKEv2 and IPsec VPNs, such as FlexVPN (SVTI-DVTI) and DMVPN~\cite{IOS-XE}. PPKs are ingested into the router from external sources using the SKIP protocol. 

The SKIP protocol operates as a restful API based on the HTTPS protocol and employs the TLS1.2 with a PSK-DHE cipher suite to ensure secure communication between the KMS and the router~\cite{IOS-XE}. Figure~\ref{fig:ppk} depicts the use-case of the Cisco SKIP protocol.
The configuration of the SKIP clients running on both the IKEv2 initiator and responder includes the IP address and port number of the key source and the preshared key for the TLS1.2 session outlined in the RFC 5246~\cite{rfc5246}. The PPK sources are set up with the SKIP parameters, comprising the local key source identity and the list of peer key source identities.
To be SKIP compliant, an external key source must implement the protocol and use an out-of-band synchronization mechanism to deliver the same PPK between the encryption devices to both the initiator and the responder. The external key source could be a QKD device, the KMS, some software, or a cloud-based key source or service.


\section{Key management solutions}
\label{sec:key-manager-solutions}

In this section, existing key management solutions are described in chronological order. Some of the solutions examined are more comprehensive than others, which only address one aspect of the key manager functionality. To the best of our knowledge, these are all publicly reported solutions related to the key management layer functionality. Although the section contains some work about the control layer, it focuses on the problems of the key management layer with the help of the central device. As a result, it is also appropriate to include and analyze them.

\subsection{DARPA quantum network key management}
\label{sec:darpa}

\begin{figure} [h]
    \centering
    \begin{subfigure}{0.55\textwidth}
        \includegraphics[width=\textwidth]{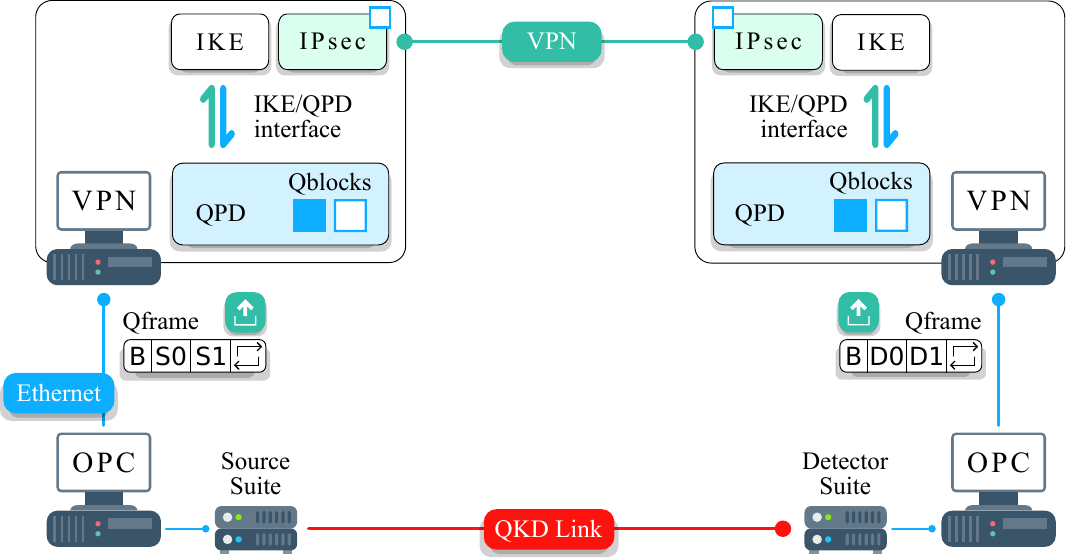}
        \caption{\label{fig:darpa-structure}}
    \end{subfigure}
    \hfill
    \begin{subfigure}{0.4 \textwidth}
        \includegraphics[width=\textwidth]{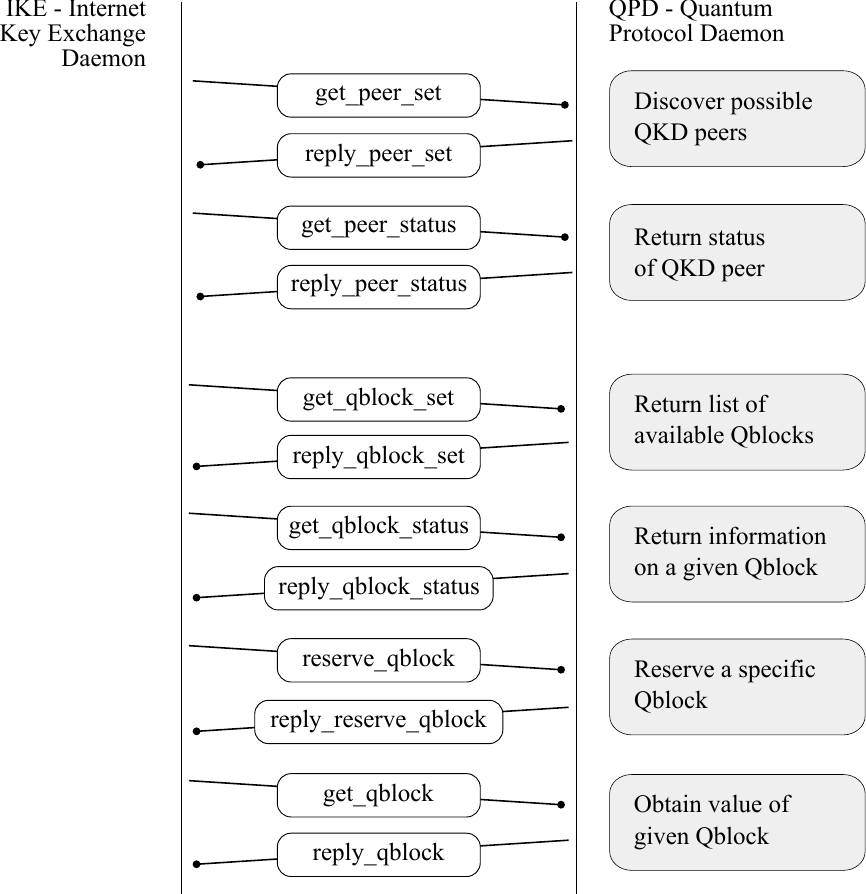}
        \caption{\label{fig:ike-qpd}}
    \end{subfigure}

    \caption{a) The DARPA quantum network structure. Secret key material is kept in fixed-sized blocks – QBlocks within QPD. QBlocks can be reserved and supplied to the IKE daemon through the IKE/QPD interface. The IKE protocol has been modified to include obtained QKD key material in the key derivation function of phase 2; b) DARPA's key supply interface, named IKE/QPD interface, allows IKE protocol, as a client application, to reserve and obtain cryptographic keys from the QKD platform.}
    \label{fig:darpa-sub}
\end{figure}

Since the DARPA quantum network was the first built QKD network, the responsible project team was the first to encounter difficulties in increasing the level of practical applicability of quantum technologies in modern communication systems~\cite{elliott2002building, elliott2003quantum, elliott2007darpa}.
As shown in Figure~\ref{fig:darpa-structure}, a QKD link connects two QKD endpoints, each comprising an optical process control (OPC) computer and a virtual private network (VPN) computer. The OPC computer oversees the optical and electronic components of the source or detector suite, facilitating the quantum transmission. The outcome, i.e., the raw key, is then forwarded to the VPN computer in a continuous series of frames known as Qframes. 

A quantum protocol daemon (QPD) running on the VPN computer distills the secret keys from the raw Qframes and stores them in memory as fixed-sized blocks known as Qblocks. 
The Qblocks can be reserved or obtained by cryptographic applications using the IKE/QPD interface, depicted in Figure~\ref{fig:ike-qpd}, where they are deployed within a key derivation function in phase 2 of the IKE protocol. Keys are always served in Qblock units of fixed size. The served Qblock is removed from QPD’s memory. While limited in functionality, this is the first practical application of QKD-derived key material, demonstrating the need to store key material and propose an interface to access QKD network services. The general concepts behind the global key distribution process are outlined without delving into greater details, but the notion of managing global keys is lacking.

\subsection{SECOQC QKD network key management}
\label{sec:sqcoqc}

The SECOQC project aimed to develop a global network for SEcure COmmunication based on Quantum Cryptography and has resulted in the first European QKD network~\cite{dianati2007transport, dianati2007architecture, dianati2008architecture, peev2009secoqc, maurhart2010qkd}. For the first time, it is made clear that the sole purpose of the QKD network is to generate, manage, and distribute ITS keys. This network is distinct from traditional telecommunication networks, with a completely new protocol stack running through all layers. It operates on dedicated network infrastructure or as an overlay network on conventional networks. The protocol stack is inspired by the traditional OSI\footnote{Open Systems Interconnection} model, but with only minor adjustments on each layer. 

\begin{figure} [h]
    \centering
    \begin{subfigure}{0.35 \textwidth}
        \includegraphics[width=\textwidth]{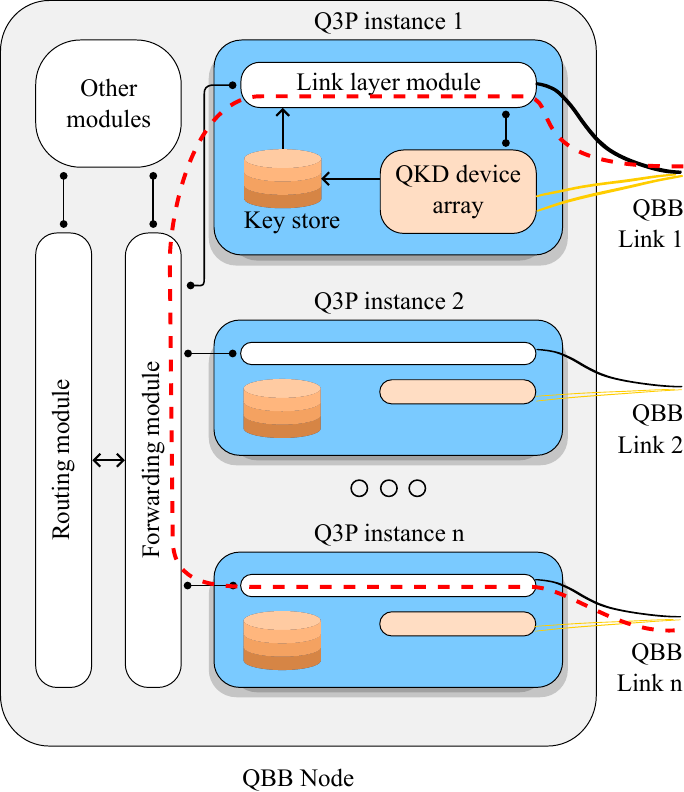}
        \caption{\label{fig:secoqc-node}}
    \end{subfigure}
    \hfill
    \begin{subfigure}{0.6 \textwidth}
        \includegraphics[width=\textwidth]{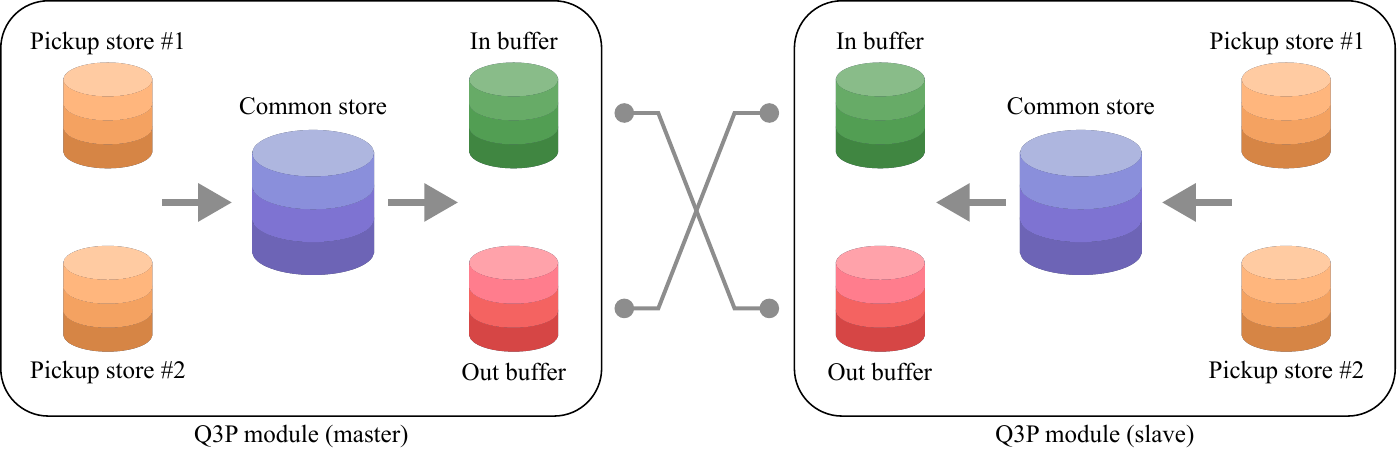}
        \caption{\label{fig:secoqc-stores}}
    \end{subfigure}
    
    \caption{a) The structure of the QBB node. The QBB node has an equal number of Q3P instances as the number of QBB links. The QBB links are logical links comprising one or more quantum channels (yellow solid lines) and a classical channel (black solid line). The dashed red line shows the traffic flow (e.g., key relay) through the QBB node. The protected packets are first processed (e.g., authenticated and decrypted) in the ingress Q3P instance. The packets are inspected for the forwarding decision and handed to the appropriate egress Q3P instance. The egress Q3P instance applies desired security profiles to the ongoing packets and forwards them over a classical IP network toward the peer QBB node; b) Key stores within SECOQC Q3P module. \textit{Pickup} stores are unique to each QKD device. When the STORE sub-protocol is completed, keys from the \textit{pickup} stores are moved to the \textit{common} store. After completing the LOAD sub-protocol, keys from the \textit{common} store are moved to \textit{in/out buffers}. The master Q3P module dictates the execution of sub-protocols. The figure illustrates the relationship between \textit{in/out buffers} between connecting Q3P modules. The \textit{in buffer} of the master Q3P module is in sync with the \textit{out buffer} of the slave Q3P module, and vice versa.}
    \label{fig:secoqc-node-sub}
\end{figure}

SECOQC's QKD network consists of Quantum Back-Bone (QBB) nodes and QBB links. The QBB node structure is depicted in Figure~\ref{fig:secoqc-node}. The Quantum Point-to-Point Protocol (Q3P), an extension of the traditional Point-to-Point Protocol (PPP), enables two points to communicate using ITS perks. This is accomplished by utilizing ITS local keys, which are stored and managed within this module. It functions as a communication interface that applies various security profiles to ongoing traffic. 

\begin{figure} [h]
    \centering
    \begin{subfigure}{0.4\textwidth}
        \includegraphics[width=\textwidth]{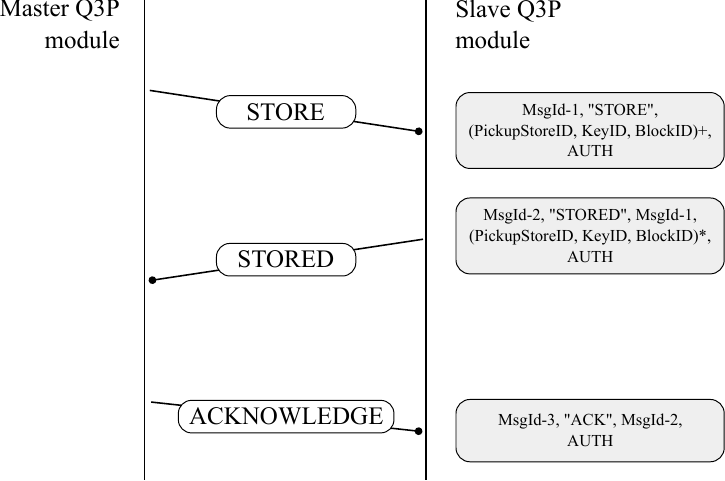}
        \caption{\label{fig:secoqc-store}}
    \end{subfigure}
    \hfill
    \begin{subfigure}{0.4 \textwidth}
        \includegraphics[width=\textwidth]{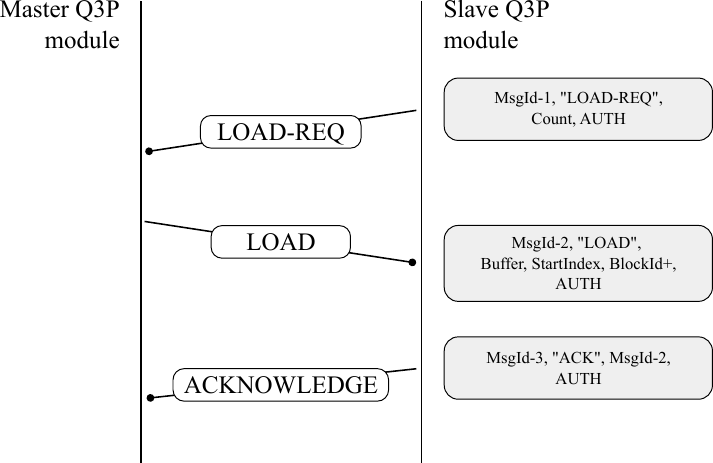}
        \caption{\label{fig:secoqc-load}}
    \end{subfigure}
    
    \caption{a) A STORE sub-protocol. It is initiated periodically by the master Q3P to transfer keys accumulated in \textit{pickup} stores to \textit{common} store. Each message is authenticated; b) A LOAD sub-protocols. It is initiated by the master with the "LOAD" message or by the slave Q3P with the "LOAD-REQUEST" message. It is used to transfer keys from \textit{common} store to a target \textit{in/out buffers} in a synchronized manner. Each message is authenticated.}
    \label{fig:secoqc-protocol-sub}
\end{figure}

When the key is generated, it is pushed from the QKD device to the Q3P module. A communication interface between QKD devices and Q3P modules must be defined to achieve interoperability. However, the interface specification is not provided or discussed further. Delivered keys are gathered in \textit{pickup} stores at the Q3P module that are specific to each QKD device (see Figure~\ref{fig:secoqc-stores}). Each key has metadata assigned to it, the most important of which is an identifier -- KeyID. Before periodically moving keys to the permanent and secure \textit{common} store, the Q3P module must ensure that the same keys are present on the peer side. A STORE sub-protocol has been proposed for this purpose. It is carried out in three stages, as shown in Figure~\ref{fig:secoqc-store}. A pair of linked Q3P modules operate in a master/slave paradigm, with the master Q3P initiating the STORE sub-protocol. 

To use key material from the \textit{common} store, the Q3P module must define key buffers, which are \textit{in-out buffers} filled with keys from the \textit{common} store and have a defined purpose. Keys from the \textit{out buffers} are only used for outbound traffic to apply security services, and keys from the \textit{in buffers} are only used for inbound traffic to process data. Both the master and the slave monitor the state of the \textit{in buffers} and can request that new keys be moved from the \textit{common} store, but only the master can decide and propose which keys should be moved.
A LOAD sub-protocol has been proposed for this purpose. It is carried out in three stages, as shown in Figure~\ref{fig:secoqc-load}. Because the LOAD sub-protocol is only triggered by receiver decisions, i.e., the state of \textit{in buffers}, Q3P modules can control the transmission rate.

The cryptographic applications establish a TCP connection with the QBB node, expecting services from the QKD network. As a result, the QBB node establishes a QKD Transport Layer (QKDTL) connection with the peer QBB node over the QKD network. The QKDTL protocol adapts the TCP protocol for the QKD network. The QKDTL handshake includes an expected key rate, and it travels hop by hop towards the destination. The SYN or SYN-ACK packet is dropped if an intermediate node cannot meet the defined key relaying throughput.\footnote{QKDTL communication is bidirectional, with both the sender and the receiver announcing desired key rates. As a result, the key resources for two different flows are probed by SYN and SYN-ACK packets.} After establishing a QKDTL connection, a random key is generated at the desired rate and relayed through the QKD network. 
Unlike the traditional TCP protocol, the QKDTL protocol does not support resending. This is why congestion control is implemented differently, specifically to respond proactively rather than reactively. If the intermediate node notices that the supply of keys is running low, it will set the CON flag within the QKDTL packet. When such a packet arrives at the destination, the destination will prolong sending the ACK, resulting in a timeout on the sender side and thus halving the congestion window. This reduces the transmission rate of the sender. 
Global keys do not need special treatment at the key management layer because they are immediately supplied to cryptographic applications.

Scheduling and load balancing are two additional concepts that can be considered within key management domain. 
The forwarding module monitors and balance the consumption load across multiple Q3P modules within the node. It is supported by the routing module which provides multiple paths in ascending order of weights, equal to the number of Q3P modules. 
If the load on the Q3P module corresponding to the shortest path exceeds some threshold value, the Q3P module on the ascending shortest path is chosen. 
Furthermore, the packets are queued for scheduling once the forwarding decision is made and the Q3P module is chosen.  

\subsection{NIST quantum network manager}
\label{sec:nist}

\begin{figure} [h]
    \centering
    \begin{subfigure}{0.44\textwidth}
        \includegraphics[width=\textwidth]{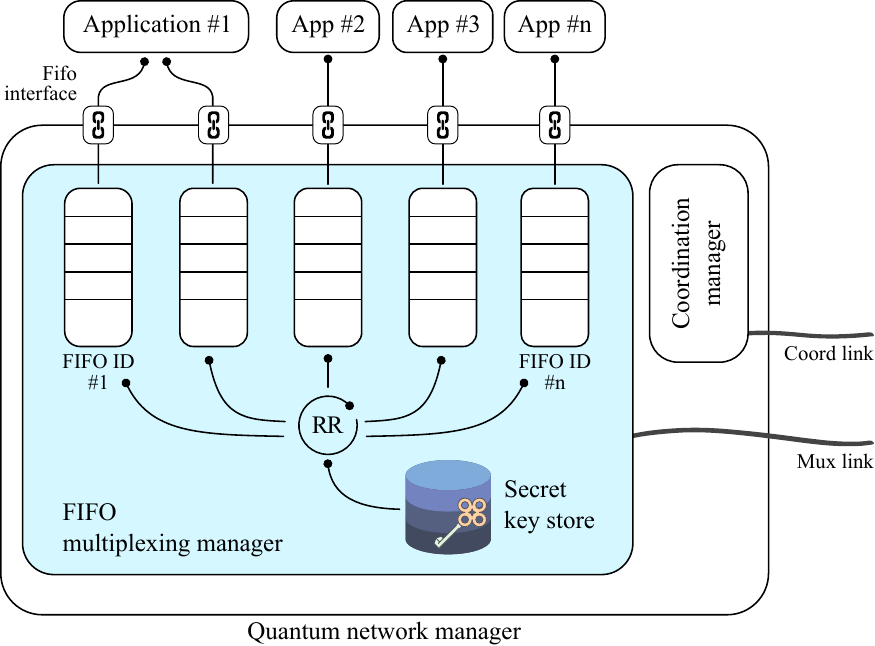}
        \caption{\label{fig:qnetwork-manager}}
    \end{subfigure}
    \hfill
    \begin{subfigure}{0.44 \textwidth}
        \includegraphics[width=\textwidth]{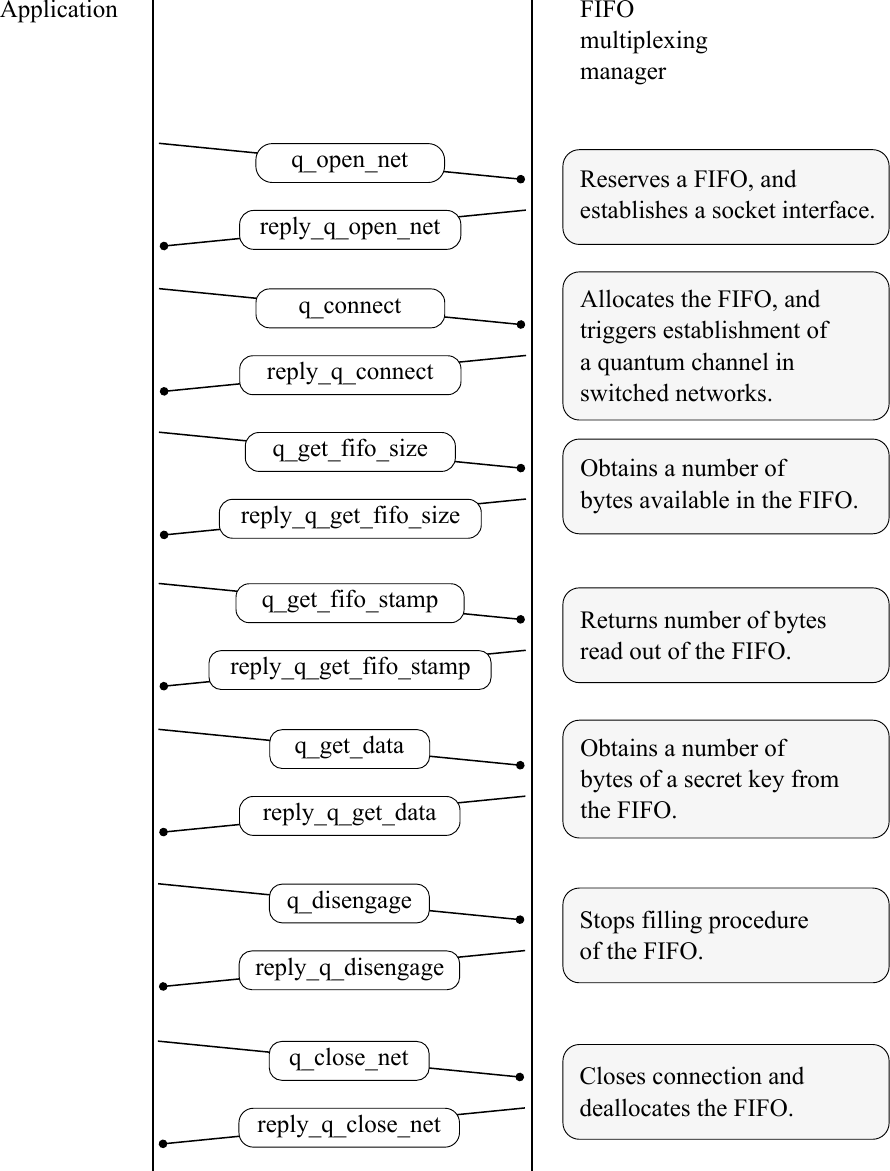}
        \caption{\label{fig:qnetwork-manager-api}}
    \end{subfigure}
    
    \caption{a) The structure of the NIST quantum network manager. The coordination manager is in charge of switching, polarization recovery and compensation, and channel timing alignment. The FIFO multiplexing manager creates and maintains an independent key stream for each connected application. It spawns the FIFO interface for each FIFO queue so that the application can access the stream of bits; b) Application programmable interface functions to define communication between applications and NIST's FIFO multiplexing manager.}
    \label{fig:nist-sub}
\end{figure}

The structure of a quantum network manager proposed in a 2008 conference paper~\cite{mink2008quantum} is depicted in Figure~\ref{fig:qnetwork-manager}. The coordination manager is responsible for control tasks and is not further discussed. The FIFO multiplexing manager creates an independent FIFO queue for a stream of synchronized bits for each application connection. A single application may open multiple threads; for example, a pair of applications may open two pairs of FIFOs for bidirectional traffic flow. Using the round-robin algorithm, the multiplexing manager fills the queues with keys from the QKD secret key store. The amount of keys assigned to each queue is set at the start of each pass. In this manner, the multiplexing manager serves multiple applications (e.g., IPsec, Transport Layer Security (TLS)) that may have different requirements (e.g., OTP, AES).

The paper~\cite{mink2008quantum} emphasizes keeping secret key stores and queues in sync across sites. A few corrupted bits do not have catastrophic consequences because a single corrupted key can be dropped and a new one used instead. On the contrary, a few dropped bits result in the loss of synchronization and the inability to use any further keys in the store or queue. This is due to the manner in which key material is stored. Each byte of generated key material is assigned a unique sequential value, a stamped ID, at the privacy amplification layer. The key material and the stamp ID of the first byte are then transferred to the secure key store of a FIFO multiplexing manager. The multiplexing manager can predict the stamp ID of an incoming material based on the most recently received entries. If it deviates from the expected value, the multiplexing manager is restarted, or it looks for the last synchronization point to preserve key material. The keys are stored in the same homogeneous manner within the FIFO queue, where each byte is now assigned a new sequential stamp ID for each queue independently. Figure~\ref{fig:qnetwork-manager-api} depicts an interface that defines communication between the application and the multiplexing manager.

\subsection{QCC security processor key manager}
\label{sec:qcc}
Within the QCC security processor, a session-based key buffer approach was proposed in 2008 to supplement the IKE/IPsec framework with QKD keys~\cite{lorunser2008security, neppach2008key}. The key manager is in charge of distributing generated key material into key buffers established for each application. The extended IKE protocol, i.e., the client application, negotiates key rates and key buffer capacities. The client first registers with a key manager by submitting its application ID and the ID of a peer application. The client then sends a connection request specifying the rate and size of the key buffer. Key managers operate on a master/slave basis, with the master having the lower IP address.

A refill procedure is initiated when the amount of key material in one of the buffers falls below a certain threshold. The threshold value can be calculated using the Equation~\ref{eq:qcc}, where $len_{max}$ is the maximum size of key requested, $size$ is the buffer capacity, and $global_{thr}$ is some globally set threshold variable. 
\begin{equation}
buff_{threshold} = max(len_{max}, size \cdot global_{thr})
\label{eq:qcc}
\end{equation}
Unassigned key material is distributed to key buffers with levels less than the threshold value during the refill procedure. The key material is distributed following application key rates. Additional key material, if any, is assigned to the remaining active key buffers. This action must be coordinated among key managers, but it is not described in detail. The communication between key managers is authenticated with QKD keys and is carried out using a binary TCP/IP protocol. 

\subsection{NEC key management}
\label{sec:nec}

\begin{figure} [h]
    \centering
    \begin{subfigure}{0.45\textwidth}
        \includegraphics[width=\textwidth]{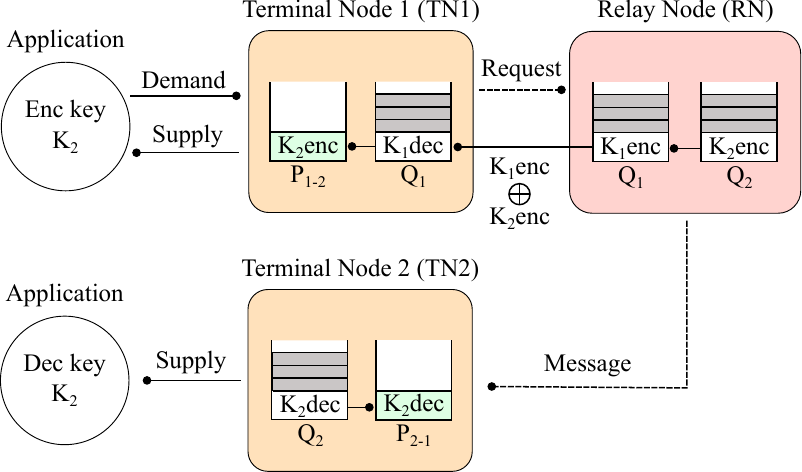}
        \caption{\label{fig:nec}}
    \end{subfigure}
    \hfill
    \begin{subfigure}{0.35 \textwidth}
        \includegraphics[width=\textwidth]{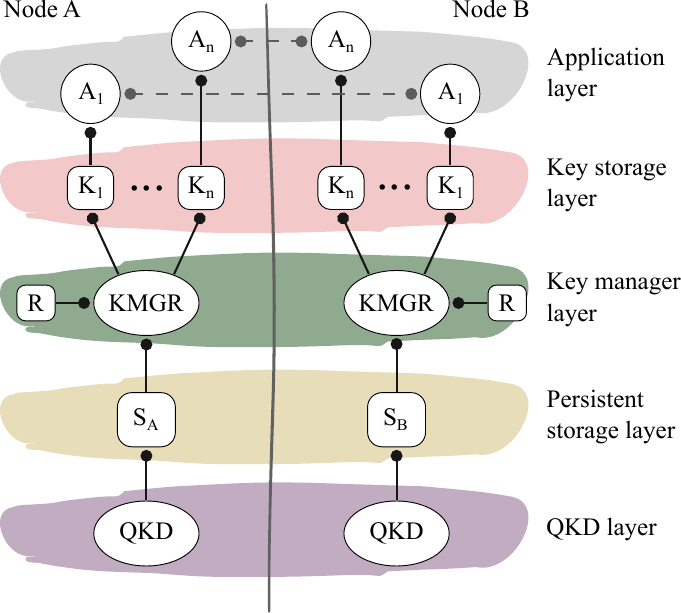}
        \caption{\label{fig:mt}}
    \end{subfigure}
    
    \caption{a) The NEC on-demand key management. On request from TN1, RN encrypts key K$_2$ using key $K_1$ and sends it to TN1. Simultaneously, RN requested that node TN2 set key $K_2$ as a global decryption key shared with TN1. TN1 decrypts key $K_2$ using local key $K_1$, and sets it as a global encryption key shared with TN2. In this manner, TN1 and TN2 share symmetric key $K_2$ defined for different purposes, and can supply requesting applications; b) Magiq Technologies layered structure.}
    \label{fig:nec-mt-sub}
\end{figure}

In 2009, authors from the NEC Corporation published a paper~\cite{maeda2009technologies} on technologies for QKD networks integrated with optical communication networks and discussed key management. The NEC architecture of the QKD network consists of four layers: a key generation layer, a connection layer, a key management layer, and a communication layer. The key generation layer is a collection of point-to-point QKD links that generate local keys. 
The connection layer performs the global key distribution. 
The key management layer monitors and controls the generation of local keys and the distribution of global keys. This can be accomplished in two ways: on-demand key supply and fixed key allocation. Figure~\ref{fig:nec} depicts on-demand key management, in which keys are relayed in response to an application request. There are two types of nodes: terminal nodes (TN) that serve applications and relay nodes (RN) that act as trusted repeater nodes and distribute keys on behalf of others. Both nodes have quantum key pools ($Q$) where local keys are stored. Terminal nodes have logical key pools ($P$) where global keys are stored. Keys are stored in fixed-size key files with one of two extensions: \textit{enc} and \textit{dec}. Local keys are always stored with \textit{dec} extension within TNs and with \textit{enc} extension within RNs. Figure~\ref{fig:nec} depicts the global key distribution. 

In the second key management technique, the fixed key allocation, the relay node creates keys in advance for all terminal nodes. This approach leads to cumbersome key management in networks with many terminal nodes, as monitoring and keeping all key pools active becomes increasingly difficult. Furthermore, supporting the addition and removal of terminal nodes from the network becomes more difficult. The fixed key allocation key management facilitates key separation in terminals and relay nodes for encryption and decryption keys.

\subsection{Magiq Technologies key manager}
\label{sec:magiq}
In 2010, a patent on key managers for QKD networks was published~\cite{keun2010key}, and we refer to the techniques proposed as Magiq Technologies (MT) key managers as the company is an assignee to the patent. MT introduces the following layers, as shown in Figure~\ref{fig:mt}: a QKD layer, a persistent storage layer, a key manager layer, a key storage layer, and an application layer. Keys are generated within the QKD layer and assigned a unique identifier as a counter value before being pushed to the persistent storage layer. Keys are kept in persistent key storage $S$ in chronological order. The node has as many persistent storages as the number of peers that are directly connected to it. The key manager layer keeps the application registration record $R$, which contains a list of connected applications $A_1, A_2,..., A_n$, references to their dedicated key storages $K_1, K_2,..., K_n$, and key rates $r_1, r_2,..., r_n$. The keys are distributed from persistent storage to application key storages based on registered application key rates. Key managers must communicate to exchange information about which keys (based on key identifiers) are distributed to which $K_i$. This is not discussed in detail, but it is known that communication is carried out using the TCP/IP protocol. The registered application then gains access to key storages to obtain keys, which are then removed from the key storage. The QKD devices can also register with the key manager to obtain the keying material required for authentication. 

The key manager may reconfigure a registration record because application key rates can fluctuate over time. The key manager monitors application key rates and adapts the distribution function dynamically to address changes. This must be coordinated among connected key managers and necessitates communication using a variant of the two-phase commit protocol. In this case, one node must act as the coordinator or the master node. Key managers may support policies that mandate all key records older than a set timestamp be deleted from the storages using the same two-phase commit protocol. 
The MT key manager includes audit and recovery functions for detecting and recovering damaged storages. 

\subsection{SwissQuantum QKD network key management}
\label{sec:swiss}

The SwissQuantum QKD network ran for more than one and a half years (from the end of March 2009 to the beginning of January 2011) to prove the long-term reliability of quantum technology, and the results were published in 2011~\cite{stucki2011long}. The network has a three-layered structure with a quantum, key management, and application layer. Although the focus is on the quantum layer, few details describing the key management layer have been revealed. At the key management layer, the key server collects keys from QKD devices, stores them, and distributes them to applications. It includes a key redundancy concept in which two different paths (direct and via trusted relay) are used to generate keys between two sites. The keys are stored in the buffers once combined with a key shared via Public Key Infrastructure (PKI) using the OTP cipher. This concept, known as dual-key agreement, is intended to improve service robustness by supplying only PKI keys when the QKD service fails. Each connected application has a separate key buffer. The application access to the buffer, i.e., the key supply interface, is not discussed in any way. Furthermore, the key relay is accomplished by sending a random key over point-to-point OTP secure tunnels. 

\subsection{QoS-supported key manager}
\label{sec:qos}

A service model and a supportive QoS-supported scheme are proposed for QKD in 2011~\cite{cheng2011qos}. There are three service classes: key-guaranteed service, key-prioritized service, and key-best-effort service. The primary performance metric used to differentiate classes is Distribution Time. The Distribution Time is the total processing time required for a global key to travel from a source node to a destination node. Based on this, the key-guaranteed service refers to applications with the greatest demand for Distribution Time and the greatest right to occupy keys. The key-prioritized service refers to applications with flexible requirements on delay, while the key-best-effort service refers to delay-insensitive applications. A Quantum Key Reservation Approach (QKRA) scheme is proposed to support key-guaranteed service class. It reserves key resources on the intermediate nodes on a relay path towards the destination. 
The hop-by-hop queue approach (HHQA) handles key prioritizing and key best-effort services. These two service classes are assigned to separate queues, with the queue serving key-prioritizing traffic receiving priority and being served first.  
The simulation results prove the advantage of the scheme when compared to QKD network without QoS support. The protocols and details of implementation are not discussed further. 

\subsection{NECTEC key management}
\label{sec:nectec}

\begin{figure} [h]
    \centering
    \begin{subfigure}{0.35 \textwidth}
        \includegraphics[width=\textwidth]{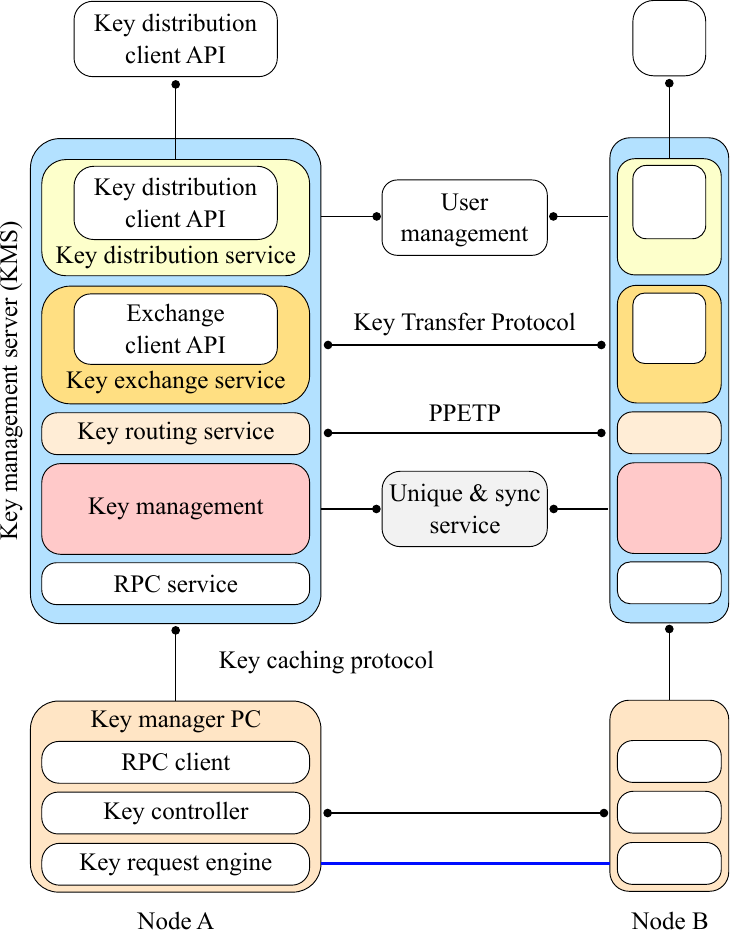}
        \caption{\label{fig:nectec}}
    \end{subfigure}
    \hfill
    \begin{subfigure}{0.4 \textwidth}
        \includegraphics[width=\textwidth]{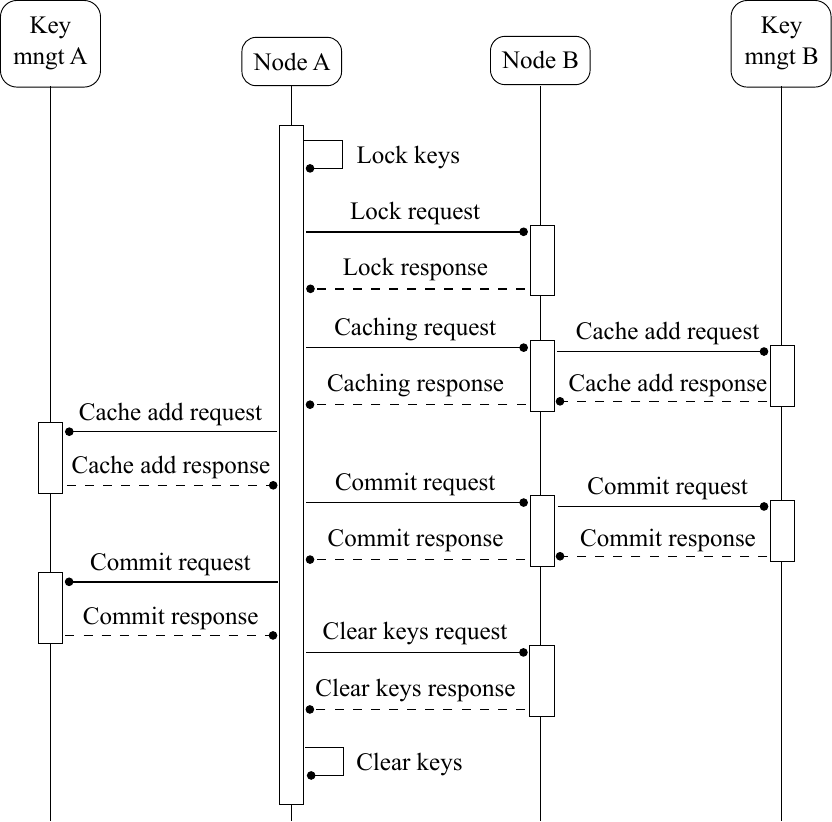}
        \caption{\label{fig:nectec-protocol}}
    \end{subfigure}
    
    \caption{a) Structure, functional elements, and protocol of the NECTEC key management approach; b) The NECTEC Key Caching Protocol defines key delivery from KMPC, where generated keys accumulate, to KMS. Before keys are delivered to KMS, they are verified.}
    \label{fig:nectec-sub}
\end{figure}

An efficient key management method has been proposed for the Thailand quantum network in studies~\cite{pattaranantakul2012secure} and~\cite{pattaranantakul2015efficient} from 2012 and 2015, respectively. We refer to this strategy as a NECTEC key management approach because the National Electronics and Computer Technology Center (NECTEC) provided the funding. The network structure consists of three layers: the quantum layer, the key management layer, and the application layer. 
Figure~\ref{fig:nectec} depicts elements and custom protocols at the key management layer. Establihed keys are first accumulated within the local Key Manager PC (KMPC). They are verified between connecting KMPCs before being handed over to the Key Management Server (KMS). 
Key Caching Protocol handles key verification and delivery to KMS. Figure~\ref{fig:nectec-protocol} depicts its sequence diagram.

The KMS uses the Key Transfer Protocol and the Point-to-Point Encrypted Transfer Protocol (PPETP) to distribute global keys. In addition to the PPETP protocol, a Key Routing protocol is defined, which determines the key distribution paths. 
A randomly generated key is encrypted with a local key shared with the next hop and sent directly to the destination. Simultaneously, the local key is relayed to the destination hop-by-hop using PPETP. A destination acquires the random key by receiving both messages. By demultiplexing a sequence of ordered secure bits into separate buffers for each application, the KMS supports multiple applications. Two buffers are distinguished depending on their purpose: an In-Buffer and an Out-Buffer. A Key Distribution Protocol is defined to supply keys to the cryptographic application but without specific details.

\subsection{Toshiba key management}
\label{sec:toshiba}

\begin{figure} [h]
    \centering
    \begin{subfigure}{0.4 \textwidth}
        \includegraphics[width=\textwidth]{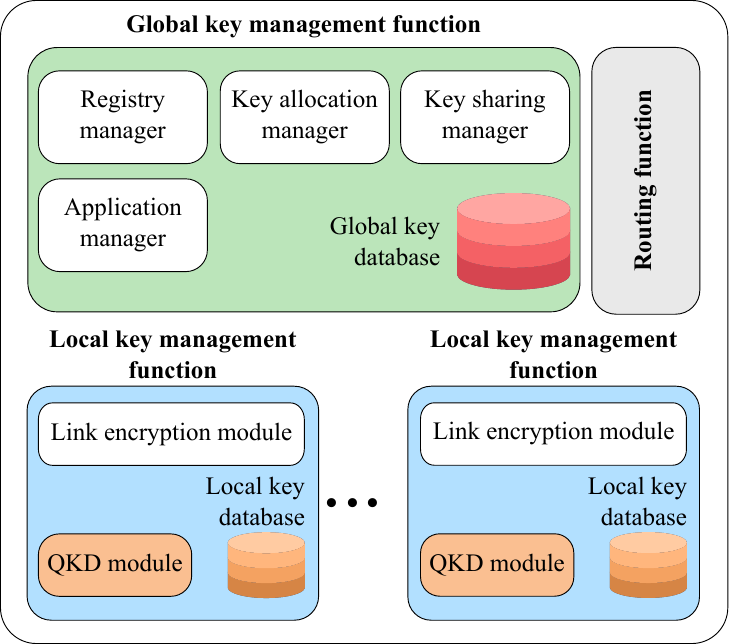}
        \caption{\label{fig:toshiba-node}}
    \end{subfigure}
    \hfill
    \begin{subfigure}{0.4 \textwidth}
        \includegraphics[width=\textwidth]{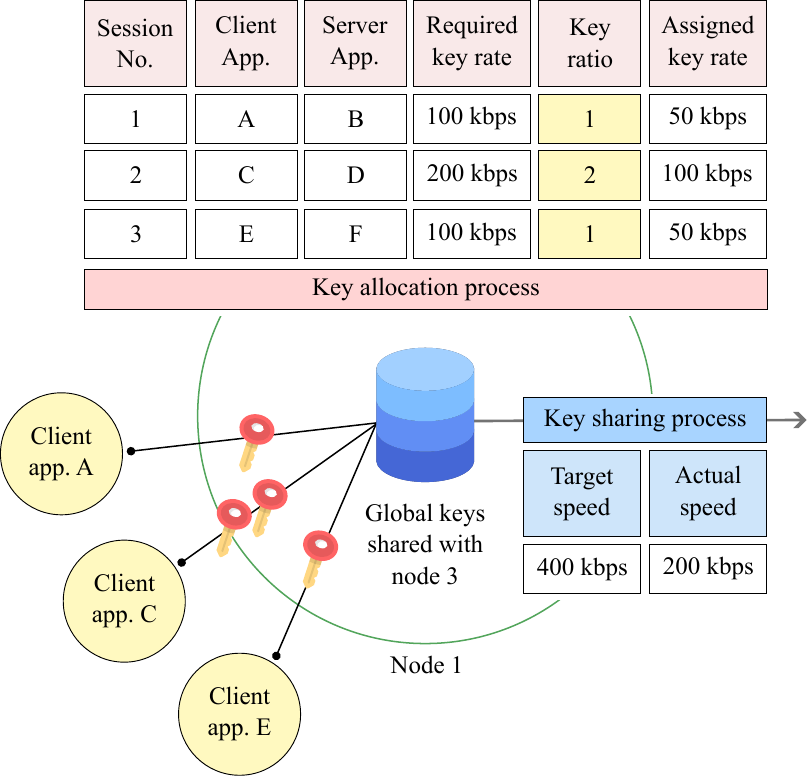}
        \caption{\label{fig:toshiba-allocation}}
    \end{subfigure}
    
    \caption{\textbf{a)} Structure of the Toshiba QKD service node. The local key management function corresponds to the SECOQC Q3P module. It manages local keys and acts as an interface for service node traffic to other nodes. The routing function selects the suitable local key management function (Q3P module) to forward keys to the next hop node. The global key management function manages global keys, including generation, distribution, and supply. This function is the main contribution of the Toshiba key management solution to the existing SECOQC approach; \textbf{b)} Key allocation process within Toshiba key management. The required rate for global key sharing (400 kbps) is the sum of the client application's required key rates (100, 200, and 100 kbps). However, the achievable global key sharing is lower (200 kbps) due to the limited generation rates of QKD links in a relay path. During the allocation process, the key ratio for connected clients is calculated and key rates are assigned accordingly. The sum of key rates assigned does not exceed the global key sharing speed. 
    }
    \label{fig:toshiba-sub}
\end{figure}

The authors from Toshiba Corporation (at the time) presented a research paper~\cite{tanizawa2016secure} in 2016 to encourage widespread use of QKD, hence the name Toshiba key management, focusing on two barriers: applicability and cost, which are usually overlooked as attention is focused elsewhere (most commonly on improving the design and implementation of QKD protocols to overcome the rate and distance limitations of technology). The presented approach was emulated, and the results show that a single QKD network can host multiple applications concurrently, fairly, and effectively. The network architecture is based on the SECOQC QKD network described in section~\ref{sec:sqcoqc} and is illustrated in Figure~\ref{fig:toshiba-node}. The local key management function corresponds to a Q3P module in the SECOQC design, and the Toshiba introduces global key management, which was partially missing in the SECOQC approach. The network architecture is enhanced with the following functions: an application directory, a key sharing and allocation mechanism, and a cryptography communication API.

The application directory function enables QKD service nodes to translate application IP addresses to the service nodes they connect.\footnote{Typically, the application key request is submitted to the local QKD node by specifying the network target application. As a result, for key relay, for example, the source node must be aware of the destination node in the entire network that serves the target application.} This function is implemented within the registry manager of a global key management function. An application directory record is created and stored on a server running on a specific service node. The remaining service nodes within the QKD network contain clients that submit and request translations from that server.  

The key sharing and allocation mechanism allows for a fair and effective supply of keys to multiple applications. As a result, two functional elements are implemented within the global key management function: a key sharing manager and a key allocation manager. The key sharing manager generates random bit sequences -- global keys, which are then relayed to the proper destination node. The rate at which global keys are generated and distributed is determined by the sum of application demands for the same destination node. To avoid congestion, the global key generation rate must be adjusted to match the actual throughput of a chain of QKD links on a relay path (see Section~\ref{sec:sqcoqc}). As a result, multiple applications are competing for access to global keys. The key allocation manager calculates the required global key ratio for each application. 
An example of a key allocation rule for three competing applications is depicted in Figure~\ref{fig:toshiba-allocation}.

To extend this work, a high-speed key management method was proposed in 2019~\cite{takahashi2019high}. To support high-speed global key distribution, an emphasis is placed on the OTP encryptor within the local key management functions. The encryptor is in the spotlight because it imposes additional computational and time costs such as encryption, decryption, local key reading, and key removal. To improve overall throughput, the authors proposed a key removal strategy.
In contrast to the conventional approach, which removes each key as soon as it is used, the proposed strategy does not remove keys until a certain number of keys are used. Keys are then removed in larger units—the results of the evaluation show that the system is adequate for the assumed high-speed QKD system.
In addition, it is revealed that the local key management function uses the SSH protocol to read keys from QKD devices, and a cryptography communication API incorporates a REST-based API that uses the HTTPS protocol. The REST-based API defines communication between applications and the global key management function and performs the following functions: providing key status, encryption key provisioning, and decryption key provisioning. It is thus equivalent to the well-established ETSI QKD 014 interface.

\subsection{NICT QKD Platform}
\label{sec:nict}

A project funded by the National Institute of Information and Communication Technology (NICT) resulted, in 2017, in a QKD platform (QKDPF) that supports multiple applications~\cite{tajima2017quantum}. Given that this paper builds on the previous research from 2011~\cite{sasaki2011field}, which describes the realization of the Tokyo QKD network, it is logical to begin there. Although this former paper should have been considered in the previous sections (to keep the approaches in chronological order), its cursory descriptions of the key management layer have resulted in its inclusion here with the introduction of QKDPF.
The Tokyo QKD network, reported in 2011~\cite{sasaki2011field}, consists of three layers: a quantum layer, a key management layer, and an application layer. The key management layer hosts Key Management Agents (KMAs). They are responsible for collecting keys from QKD devices, reshaping keys, assigning identifiers, and storing keys in numerical order for encryption and decryption. Once again, the interface between QKD devices from various vendors and key management is acknowledged but not discussed in detail. It is unclear how KMAs store, manage and use keys, but it is stated that user data is given to the KMA for encryption. This goes against the now-well-established definition and purpose of QKD networks: to generate and distribute ITS cryptographic keys rather than securely transfer user data. Furthermore, the term Key Management Server (KMS) refers to a centralized entity that assists KMA by performing key lifecycle management and relay path provisioning.

\begin{figure} [h]
    \centering
    \begin{subfigure}{0.45 \textwidth}
        \includegraphics[width=\textwidth]{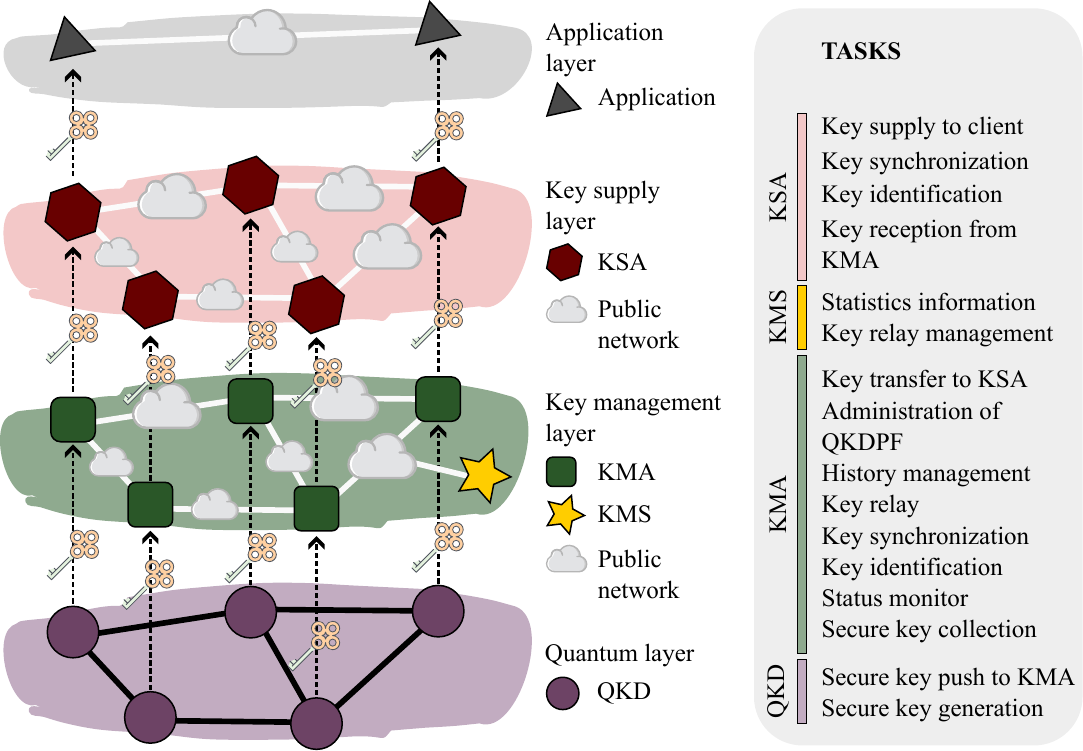}
        \caption{\label{fig:nict-structure}}
    \end{subfigure}
    \hfill
    \begin{subfigure}{0.45 \textwidth}
        \includegraphics[width=\textwidth]{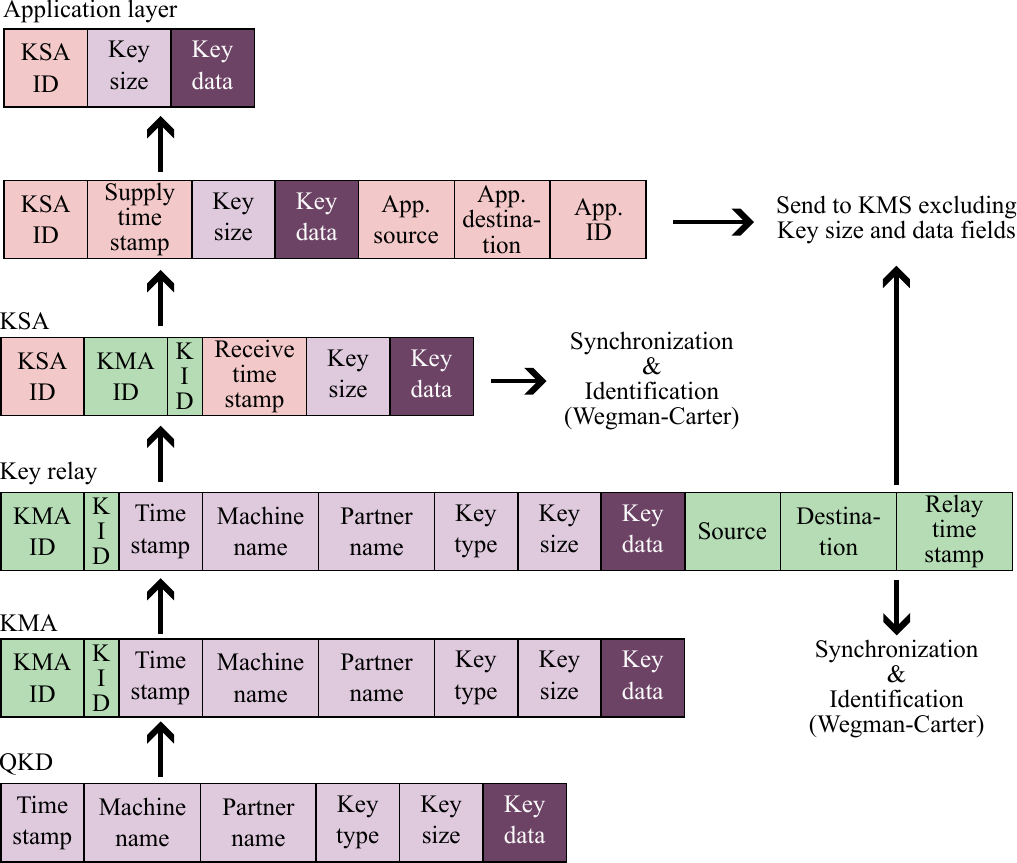}
        \caption{\label{fig:nict-lifecycle}}
    \end{subfigure}
    
    \caption{a) NICT QKD Platform structure and functional requirements of each layer; b) NICT's key lifecycle. The key format is explained as follows: \textit{timestamp} -- time stamp of quantum key generation; \textit{machine name} -- ID of QKD equipment at local site; \textit{partner name} -- ID of QKD equipment at opposite site; \textit{key type} -- identifier of encoding key or decoding key; \textit{key size} -- key size; \textit{key data} -- key data; \textit{KMA ID} -- ID of KMA; \textit{KID} -- ID of received quantum key; \textit{relay time stamp} -- time stamp of key relay; \textit{source} -- relay source; \textit{destination} -- relay destination; \textit{KSA ID} -- ID of KSA; \textit{receive time stamp} -- time stamp of key reception; \textit{supply time stamp} -- time stamp of key supply; \textit{app. source} -- source ID application; \textit{app. destination} -- destination ID application; \textit{application ID} -- application ID.}
    \label{fig:nict-sub}
\end{figure}

The QKDPF, proposed in 2017~\cite{tajima2017quantum}, extended the layered structure of the Tokyo QKD network with a key supply layer, thereby aligning it with the definition and general purpose of QKD networks. This newly introduced layer sits between the key management and application layers, as shown in Figure~\ref{fig:nict-structure}. It hosts Key Supply Agents (KSAs), enabling the secure supply of independent cryptographic keys to multiple applications. Unlike the original Tokyo QKD network design, the KMAs now transmit random data (cryptographic keys) hop by hop instead of user data. Figure~\ref{fig:nict-lifecycle} depicts the key lifecycle and general information associated with keys at various layers. The ITU-T Y series recommendations now advocate for nearly identical structure, elements, functions, and key formats to one proposed within QKDPF.

\subsection{Quantum Canada key management}
\label{sec:canada}

A project funded by Quantum Canada resulted in 2018 in general guidelines for designing a QKD network structure suitable for deployment in enterprise environments~\cite{tysowski2018engineering}.
The network structure comprises four layers: the QKD link layer, the network layer, the key management layer, and the application layer. QKD is performed on individual point-to-point links at the QKD link layer, generating keys between neighboring nodes. The keys are then passed to the network layer, which manages these local keys and allows key distribution between arbitrary network nodes. Key distribution occurs at the request of the KMS layer, which estimates the load based on end-user demand for keys.

In terms of key management, Quantum Canada's approach is broadly similar to Toshiba's approach (Section~\ref{sec:toshiba}), and thus the SECOQC project (Section~\ref{sec:sqcoqc}). The network layer manages local keys and performs key relaying. It is divided logically into two planes: control and data planes. Based on demands generated by KMS, which can be either continuous or one-time, the control plane manages global key generation between sites and determines the best relaying paths. The continuous mode requires a specific key generation rate with a remote site, whereas the one-time mode requires a particular number of keys generated with a remote site. The data plane carries out key relaying, which uses local keys from the temporary key pool management function. 
Because multiple relaying flows for distinct remote sites share key resources of a QKD link layer, a scheduling algorithm is implemented to ensure fairness. A deficit-weighted round-robin algorithm is used for continuous mode demand, and a simple FIFO queuing mechanism is used for one-time demand.
Local keys are classified into three types and thus assigned to a specific purpose. A portion of the local keys are reserved for distribution to hosts on directly connected sites (no relaying is required). Others are used for relaying purposes, and there are two types. The first type includes local keys that will be transformed into global keys as a result of relaying between the local and remote sites. The second type includes local keys, used within the local site to relay keys on behalf of others.

Keys managed by the network layer are eventually passed to the KMS layer on demand. The KMS collects, stores and maintains synchronization of keys.
Multiple connected clients within a secure site share access to the key storage. As a result, the authors emphasize the issue of key access collisions, which result in key material waste. The key access collision occurs when a client application within site A is served with key K while the KMS in remote site B serves the same key K for a different purpose. Several solutions to this problem are proposed. One solution is for KMSs to serve keys from different ends of the database, such as the KMS at site A serving keys from the beginning and the KMS at remote site B serving keys from the end. This solution, however, is not recommended because it may result in practical inefficiency due to key pool fragmentation. Another way to solve key access collisions is to assign a small number of keys from the quantum key pool to a working set and have KMSs serve keys from that working set only in a previously described manner.

Client requests are monitored to collect and analyze the demand statistics, which are used to control global key generation at the network layer. 
If the available keys are deemed insufficient, a policy engine function may provide a fall-back method, such as producing keys based on a key derivation function. The policy engine function defines policy rules governing key use and includes client expectations (size and lifetime, for example). Client expectations are classified into five categories based on security requirements, and these categories help guide the key generation process at the network layer.

\subsection{NSFC SDQaaS framework}
\label{sec:sdqaas}

This subsection summarizes the research conducted through several works based on the same concept but introducing various optimizations. The concept will be referred to as the NSFC solution because the National Natural Science Foundation of China (NSFC) provided ongoing support. First, an SDQaaS framework is introduced in 2019~\cite{cao2019sdqaas} (skipping the preliminary work from 2017~\cite{cao2017key}). SDQaaS is an acronym that stands for SDN for QKD as a service. The QKD as a service (QaaS) concept shares the QKD network infrastructure among multiple clients. Furthermore, in SDQaaS, the QaaS is implemented in a centralized SDN controller to provide efficient and flexible key allocation. The network structure consists of three planes: infrastructure, control and application planes. 

QKD nodes, placed in the infrastructure plane, are equipped with Open Flow Agents (OFA), which communicate and share relevant information with the SDN controller via the Open Flow Protocol (OFP). 
In a centralized control layer, a topology module collects and stores information about the QKD network topology and nodes, whereas a resource module deals with more dynamic data, gathering and storing real-time secret key rates of QKD links. Multiple clients request secret-key rate (SKR) settings from the QKD infrastructure via the northbound interface, which is realized as a REST-based API using the HTTP protocol. The interface provides three simple methods for creating, modifying, and deleting services. 
According to the requirements, a centralized control layer allocates available SKR on each QKD link along a path to fulfill service requests from multiple clients.

The same authors proposed the Multi-Tenant Key Algorithm (MTKA) in a 2019 study~\cite{cao2019multi}. It adheres to the same principles as SDQaaS in providing a centralized and detailed view of the QKD network infrastructure and its resources.  
Efficient secret key resource usage can be achieved by maximizing the Matching Degree (MD) function, a sum of success probability, and key resource utilization multiplied by weighting factors $\alpha$ and $\beta$. The success probability is defined as a ratio of admitted client requests to total requests, while key resource utilization is defined as a ratio of assigned SKR slots to total SKR slots. The MTKA algorithm assumes that the set of client requests, network topology, and SKRs on each QKD link are known in advance. It returns a list of admitted requests and the number of occupied key resources in a network. The MD function should then be optimized to maximize key resource usage.

\subsection{NKPs DDKA-QKDN scheme}
\label{sec:nkps}

In 2022, a Dynamic-On-Demand Key Allocation (DDKA) scheme for QKD networks (QKDN) has been proposed~\cite{chen2022ddka}. To efficiently use key resources in the Quantum Internet-of-Things (Q-IoT) scenario, the DDKA-QKDN scheme dynamically allocates key resources per the application request. The scheme simultaneously addresses the key supplement of QKPs as well. Again, the concept is realized using SDN technology.

The primary criterion used to process application requests is request arrival time. However, due to the large number of requests that can arrive simultaneously, requests are queued for processing. A new strategy that prioritizes queues had to be developed to decrease queuing delays. There are two main request requirements: request key quantity $K_{qua}$ and security $K_{sec}$. These two requirements are directly related, as the greater key quantity implies greater key security and vice versa. Security levels $Sec$ are different for different applications, ranging from low-security levels ($Sec = 0$), where data is transmitted in plaintext, to high-security levels ($Sec > 0$), where different-sized keys are used to provide data confidentiality. Keys with the following sizes are available based on security level: 128 bits, 256 bits, 512 bits, 1024 bits, and 2048 bits. Smaller key quantity requests are given higher priority regarding system efficiency when providing keys. As a result, storages are consumed slowly, and the likelihood of serving subsequent requests increases. When it comes to system security, however, higher key security requests are prioritized. As a result, when calculating response weight value $est(K_i)$ the DDKA-QKDN scheme accounts for the trade-off between the two using $\omega \in [0,1]$, as shown by the Equation~\ref{eq:nkp}. When multiple requests arrive simultaneously, they are sorted in ascending order of response weights.

\begin{equation}
est(K_i) = (1 - \omega)lnK_{qua} + \omega ln(10 - K_{sec})
\label{eq:nkp}
\end{equation}

If the request cannot be fulfilled due to a lack of keys, a request is made to supplement the pool with new keys. Key supplement requests are handled in the same manner as application key requests. The primary criterion is request arrival time, and the response weight value is used to process requests that arrive at the same time in an orderly manner. Since the key supplement process can take a long time to complete (might require a key relay process), thereby prolonging the waiting time of application requests, storage thresholds are introduced to manage key supplements dynamically. As a result, storages are promptly supplemented to prevent exhaustion.
On the other hand, a higher threshold value is set to prevent overloading storages with key material that might not be consumed anytime soon and could reduce key security.

\subsection{KISTI key management}
\label{sec:kisti}

Korea Institute of Science and Technology Information (KISTI) research presented a QKMS design plan in 2022 to assure the physical layer security of the next generation KREONET\footnote{KREONET is a national research and development network managed by KISTI.}~\cite{shim2022design, shim2022bdesign}. The architecture is layered, with transport and quantum planes decoupled. This paper only describes the high-level design of functional blocks, and there isn't much to reflect on in implementation. The structure is identical to that of the NICT QKD platform, including the quantum, KMA, and KSA layers. Every node in the network creates and keeps keys with every other node. As a result, two key pools are distinguished: those that store key material between adjacent nodes via quantum connectivity and those that store key material generated via key relay. Keys supplied from the QKD devices are intercepted in the QKD protocol abstraction layer and converted to the ETSI QKD 014 standard format before being delivered to KMA.
Like Quantum Canada's approach, the scheme defines fallback methods to deal with scarce key resources efficiently. The first option provides key relay even when the key material available on the link is insufficient to meet the request. The encryption key is derived from the quantum key using HKDF, which is then used to OTP encrypt the relay key. The second method employs HKDF on the quantum key to generate supply keys. The authors conclude that because HKDF is used to generate supply keys, quantum keys can be kept in key pools in fixed sizes. Because the application requests different key sizes, this requirement is filled using HKDF. However, if implemented in this manner, the QKD network will be unable to serve true ITS keys. This is because classical key expansion methods do not produce true random output.

\subsection{AIT key manager}
\label{sec:ait}

In 2023, authors from the Austrian Institute of Technology (AIT) in Vienna published a report on key manager being developed within the EuroQCI framework~\cite{james2023key}. Their paper describes a KMS prototype that adheres to ETSI QKD 004, 014, and 015 standards and considers ITU-T Y series recommendations. The main contributions of AIT are recommendations for KMS-to-KMS and KMS-to-SDN agent interface methods. In addition, the ETSI QKD 004 interface has been enhanced with a push mode to support communication between QKD devices and KMS. 

\begin{figure} [h]
    \centering
    \begin{subfigure}{0.4 \textwidth}
        \includegraphics[width=\textwidth]{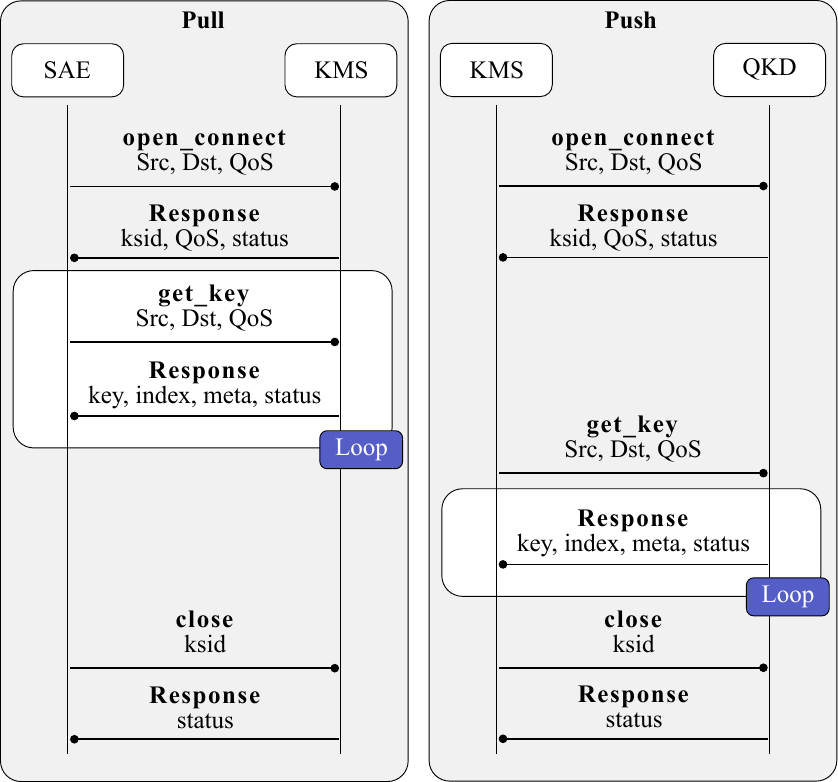}
        \caption{\label{fig:ait-kmngt}}
    \end{subfigure}
    \hfill
    \begin{subfigure}{0.4 \textwidth}
        \includegraphics[width=\textwidth]{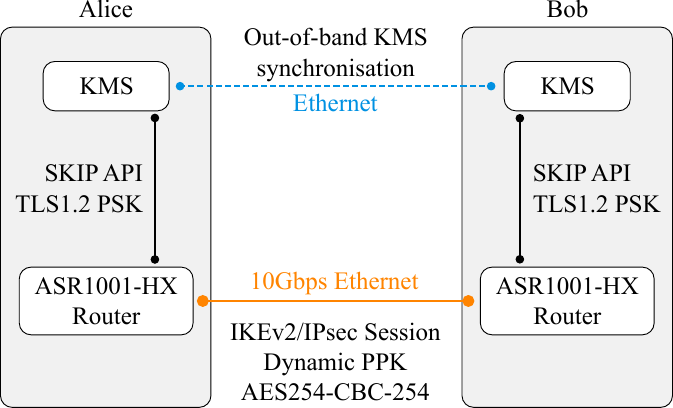}
        \caption{\label{fig:lab-setup}}
    \end{subfigure}
    
    \caption{a) The enhancement of the ETSI 004 interface with a push mode defined by the AIT. It is used to supply generated keys from the QKD devices to the KMS; b) Cisco Lab setup.}
    \label{fig:ait-cisco-sub}
\end{figure}

The ETSI QKD 004 interface specification advocates using an interface between KMSs of different hierarchical levels within the same node and between QKD devices and KMS. However, the authors highlight differences between the key-supply interface and the QKD device-KMS interfaces. In general, applications want to pull keys from the KMS on demand, whereas QKD devices wish to push keys to the KMS once generated. The authors propose a push mode for the ETSI QKD 004 interface for these reasons. Figure~\ref{fig:ait-kmngt} illustrates the pull and push mode.

AIT key manager prototype proposes methods for key modifications, applications, key streams, and peer availability and finally defines a concrete protocol to carry these methods. The protocol of choice is a CoAP protocol. It is a REST-based, specialized web transfer protocol designed for limited-capability devices. For key modification, the following methods are introduced: 
\begin{itemize}
    \item \textbf{new\_key\_batch} -- synchronizes key material obtained from QKD devices. To reduce internal key consumption, keys are synchronized in batches. The message contains key IDs and a Message Authentication Code (MAC) calculated from the message and key data transmitted.
    \item \textbf{forward\_keys} -- performs key relay and includes encrypted keys, key IDs, and a destination ID that guides subsequent hops.
    \item \textbf{split\_key} -- splits key in smaller key blocks. It contains a key ID, a list of new key IDs, and their corresponding new lengths.
    \item \textbf{merge\_keys} -- merges smaller keys into one larger key. It contains a list of key IDs, a new key ID, and the corresponding new length.
    \item \textbf{delete\_keys} -- deletes keys identified with key IDs.
    \item \textbf{make\_keys\_internal} -- reserves a group of key identified with given key IDs for internal use. This includes authentication and encryption keys for synchronization and relay purposes.
    \item  \textbf{make\_keys\_external} -- reserves a group of key identified with given key IDs for peer applications. It contains a list of key IDs and a key stream ID. 
\end{itemize}

When registering a new application key stream, a peer KMS is notified using the \textbf{new\_app} method. This method includes an application ID, the source and destination address, a key stream ID, and a QoS. Closing of the application key streams is synchronized using \textbf{key\_stream\_closed} notification. It includes a key stream ID. The KMSs can obtain or inform peers about their status using \textbf{get\_status} and \textbf{post\_status} methods. An interface is also defined between the KMS and the SDN agent, located within the same security boundary. 
In addition, the authors discuss PQC hybridization techniques at the KMS level. This is how the dual secret key agreement is carried out.

\subsection{CISCO Key Management System}
\label{sec:cisco}
One of the Cisco testbeds is illustrated in Figure~\ref{fig:lab-setup} and consists of the KMS and a Cisco Router ASR1001-HX. The KMS collects keys from the various vendor's QKD devices and makes them accessible to the network routers through the SKIP protocol. Additional experiments were carried out to transfer keys over complex and arbitrary topologies using trusted-repeater nodes.
The KMS is constructed utilizing the Python Flask framework as its API gateway. A proxy server fronts the KMS to safeguard the API gateway from possible attack vectors~\cite{IOS-XE}. Further security is provided through the implementation of the TLS1.2 PSK authentication between the routers and the KMS.

The KMS acquires keys in bulk by reading binary files from the QKD device. The sequence of file retrieval varies depending on the QKD vendor's specifications. The binary file is converted into a hex format at the KMS to minimize storage requirements. The hex-formatted keys are stored in 64-bit blocks as ASCII text in the internal SQLite database. The availability of keys in the database is continuously monitored, and when a certain threshold is reached, a process is initiated to replenish the database on time.

The keys are supplied to the routers through the Cisco SKIP protocol. As previously mentioned, this connection is secured with the TLS to ensure client authenticity. 
Each router first registers itself using a unique ID to ensure traceability and identification of local and remote devices. The first requester is expected to initiate a key exchange with the remote key storage device. The two key storages will jointly formulate a Key and a KeyID. The second GET request from the remote system is made after the Key Storage has been negotiated with its counterpart. The key storage negotiates a key with the given KeyID and transfers its copy of the Key and KeyID to the router in the query response.


\section{Discussion}
\label{sec:discussion}
A detailed overview of the evolution of key managers is given in section~\ref{sec:key-manager-solutions} through a detailed analysis of the solutions in the chronological order of their appearance. Considering the timeline spanning from their earliest iterations aimed at achieving basic functionality to the present-day versions boasting enhanced features and optimizations, it becomes challenging to individually acknowledge each solution while giving due credit. By considering the functional requirements outlined in the ITU-T recommendations for the key management layer~\cite{itu3801}, we can assess how various solutions address these requirements, if at all. In the following subsections we consider several key requirements and functionalities of key managers and conduct a comparative analysis of existing solutions. By conducting this analysis, we aim to pinpoint the current gaps and challenges in addressing the key management problem within QKD networks.

\subsection{Compatibility with various kinds of QKD modules}
\label{sec:compatibility}
The key managers are required to collect keys from the QKD modules via appropriate interface and be compatible with various kinds of QKD modules which implement different protocols. While this requirement appears straightforward to achieve, most commercial key managers are packaged with QKD modules. As a result, they use proprietary methods of communication between QKD modules and key manager agents. To facilitate fair representation of all QKD equipment manufacturers in the market, it's imperative to establish clear guidelines for communication and interoperability between various QKD module manufacturers and key managers. The ETSI 004 application interface is designed to facilitate communication among key managers at various hierarchical levels. As a result, it is possible to keep a collection of proprietary QKD modules and key managers while installing a hierarchically superior key manager from any manufacturer that collects keys through the interface. However, the ETSI 004 application interface is not widely used in practice due to QoS features which are yet difficult to support. In addition, a QKD module manufacturer may not implement its own key manager solution. As a result, the standardized interface for delivering keys from QKD modules to key managers is critical to achieving interoperability.

This requirement was first highlighted during the SECOQC project (Section~\ref{sec:sqcoqc}), when it was critical to allow different implementations of QKD protocols to deliver keys to the Q3P modules. But the interface was never explained in detail, and its implementation is unknown (at least to the general public). Although the interface between QKD modules and key managers is frequently mentioned, it was not previously described until recently when ETSI interfaces were modified for this purpose. The KISTI key management solution (Section~\ref{sec:kisti}) addresses the issue via a QKD protocol abstraction layer. The abstraction layer collects keys generated by heterogeneous QKD modules, converts them to the ETSI 014 standard format, and delivers them to the key manager for storage. However, it is unclear how keys are transferred to the abstraction layer and what format the message takes. The AIT key manager design proposes a more comprehensive solution, which includes an extension of the ETSI 004 interface (Section~\ref{sec:ait}). It defines a push mode for the GET\_KEY method, which delivers keys in standard-defined format. This approach supports the capability to develop QKD modules independently of a key manager, as keys can be transmitted immediately after generation.

\subsection{Key supply to the user network}
\label{sec:supply-discussion}
The key manager is required to provide requested number of keys to cryptographic applications via a key supply interface with security capabilities. It is further required that key manager applies the key management policies. In this context, we examine several factors concerning access to QKD network services: the key-supply interface, the approach to resource sharing, fallback methods, and application priorities. Table~\ref{tab:key-supply} provides a summary of these features for the current approaches.

\renewcommand{\arraystretch}{1}
\begin{table}
\centering
\caption{Comparison of the key managers in terms of key-supply interface functionalities.}
\resizebox{0.98\textwidth}{!}{%
\begin{tabular}{lcccccc}
\toprule
Features                                                                            &                        &                                                                                             &                                                                                                                                                                  &                                                                                                                      &                              &                                   \\ \cline{1-1}
Solutions                                                                           & \multirow{-2}{*}{Year} & \multirow{-2}{*}{\begin{tabular}[c]{@{}c@{}}Key-Supply Interface\\ Definition\end{tabular}} & \multirow{-2}{*}{\begin{tabular}[c]{@{}c@{}}Key-Supply Interface\\ Capabilities\end{tabular}}                                                                    & \multirow{-2}{*}{Resource sharing}                                                                                   & \multirow{-2}{*}{Priorities} & \multirow{-2}{*}{Fallback method} \\ \midrule

\begin{tabular}[c]{@{}l@{}}DARPA key\\ management\end{tabular}                      & 2002-2007              & Yes                                                                                         & \begin{tabular}[c]{@{}c@{}}List available Qblocks;\\ Reserve a certain Qblock;\\ Obtain Qblock.\end{tabular}                                                     & No                                                                                                                   & No                           & No                                \\ \hline
\begin{tabular}[c]{@{}l@{}}SECOQC key\\ management\end{tabular}                     & 2007-2009              & No                                                                                          & \begin{tabular}[c]{@{}c@{}}A TCP connection with specified\\ key-supply rate.\end{tabular}                                                                       & Reservation based approach.                                                                                          & No                           & No                                \\ \hline
NIST manager                                                                        & 2008                   & Yes                                                                                         & \begin{tabular}[c]{@{}c@{}}Establish a session;\\ Get number of available bytes;\\ Obtain a number of bytes;\\ Disengage; Close the session.\end{tabular}        & \begin{tabular}[c]{@{}c@{}}Round Robin sharing algorithm\\ based on registered application\\ key rates.\end{tabular} & No                           & No                                \\ \hline
\begin{tabular}[c]{@{}l@{}}QCC security\\ processor\\ key manager\end{tabular}      & 2008                   & Yes                                                                                         & \begin{tabular}[c]{@{}c@{}}Establish a session with defined\\ key rate and key buffer size.\end{tabular}                                                         & \begin{tabular}[c]{@{}c@{}}Sharing based on registered\\ application key rates.\end{tabular}                         & No                           & No                                \\ \hline
\begin{tabular}[c]{@{}l@{}}NEC key\\ management\end{tabular}                        & 2009                   & No                                                                                          & -                                                                                                                                                                & No                                                                                                                   & No                           & No                              \\ \hline
\begin{tabular}[c]{@{}l@{}}Magiq Tehnologies\\ key manager\end{tabular}             & 2010                   & No                                                                                          & -                                                                                                                                                                & \begin{tabular}[c]{@{}c@{}}Sharing based on registered\\ application key rates.\end{tabular}                         & No                           & No                                \\ \hline
\begin{tabular}[c]{@{}l@{}}SwissQuantum\\ key management\end{tabular}               & 2009-2011              & No                                                                                          & -                                                                                                                                                                & No                                                                                                                   & No                           & Yes                               \\ \hline
\begin{tabular}[c]{@{}l@{}}QoS-supported\\ key manager\end{tabular}                 & 2011                   & No                                                                                          & -                                                                                                                                                                & \begin{tabular}[c]{@{}c@{}}Sharing based on priority class.\\ Includes reservation based approach.\end{tabular}      & Yes                          & No                                 \\ \hline
\begin{tabular}[c]{@{}l@{}}NECTEC key\\ management\end{tabular}                     & 2012                   & No                                                                                          & -                                                                                                                                                                & No                                                                                                                   & No                           & No                               \\ \hline
\begin{tabular}[c]{@{}l@{}}Toshiba\\ key management\end{tabular}                    & 2016                   & Yes                                                                                         & \begin{tabular}[c]{@{}c@{}}Providing status;\\ Encryption key provisioning;\\ Decryption key provisioning\\ (evolved in ETSI GS\\ QKD 014 standard)\end{tabular} & \begin{tabular}[c]{@{}c@{}}Sharing based on registered\\ application key rates.\end{tabular}                         & No                           & No                               \\ \hline
\begin{tabular}[c]{@{}l@{}}NICT QKD\\ platform\end{tabular} & 2017                   & No                                                                                          & -                                                                                                                                                                & No                                                                                                                   & No                           & No                               \\ \hline
\begin{tabular}[c]{@{}l@{}}Quantum Canada\\ key management\end{tabular}             & 2018                   & No                                                                                          & -                                                                                                                                                                & No                                                                                                                   & Yes                          & Yes                               \\ \hline
\begin{tabular}[c]{@{}l@{}}NSFC SDQaaS\\ framework\end{tabular}                     & 2019                   & No                                                                                          & -                                                                                                                                                                & Reservation based approach.                                                                                          & No                           & No                               \\ \hline
\begin{tabular}[c]{@{}l@{}}NKPs\\ DDKA-QKDN\\ scheme\end{tabular}                   & 2022                   & No                                                                                          & -                                                                                                                                                                & No                                                                                                                   & Yes                          & No                               \\ \hline
\begin{tabular}[c]{@{}l@{}}KISTI\\ key management\end{tabular}                      & 2022                   & Yes                                                                                         & ETSI GS QKD 014                                                                                                                                                  & No                                                                                                                   & No                           & Yes                               \\ \hline
\begin{tabular}[c]{@{}l@{}}AIT\\ key manager\end{tabular}                           & 2023                   & Yes                                                                                         & \begin{tabular}[c]{@{}c@{}}ETSI GS QKD 014;\\ ETSI GS QKD 004\end{tabular}                                                                                       & No                                                                                                                   & No                           & No                                \\ \hline
\begin{tabular}[c]{@{}l@{}}Cisco\\ key manager\end{tabular}                           & 2023                   & Yes                                                                                         & SKIP protocol                                                                                      & No                                                                                                                   & No                           & No                                \\ \bottomrule
\end{tabular}
}
\label{tab:key-supply}
\end{table}
\renewcommand{\arraystretch}{1}

While two standard interfaces have been established to regulate the communication between cryptographic applications and key managers, it's still valuable to analyze earlier proposed interfaces for comprehensive understanding. The earliest form of this interface was suggested within the DARPA quantum network framework (Section~\ref{sec:darpa}). Cryptographic applications can reserve and retrieve keys from the QKD network using this straightforward interface. However, it is the applications' responsibility to negotiate and reserve keys before they can be retrieved. The interface does not allow for any additional requirements, such as requested key sizes. 
The functionalities and approach of this interface are outdated, rendering it unsuitable for application in today's implementations. After several years Toshiba defined a similar interface, which evolved into the current ETSI QKD 014 standard. It is used in the newer KISTI and AIT key management solutions. The NIST and QCC interfaces can be linked to ETSI QKD 004 key session establishment logic. Both require the establishment of key sessions with defined application key rates. 
However, these implementations lack support for QoS, which is why the ETSI QKD 004 standard exists. Both approaches ensure a fair sharing of resources, but the supply of keys according to the application's QoS requirements is not guaranteed. Finally, Cisco defined its key-supply interface, the SKIP protocol. However, according to the publicly available description provided (see section~\ref{sec:skip}), the underlying methods of the protocol are very similar to the ETSI QKD 014. From this discussion, it's evident that two prevailing approaches to implementing the interface exist: one is simplistic, enabling key retrieval on demand, while the other entails establishing a session with desirable key rate. The ETSI interfaces effectively encompass the functionalities of all previous approaches.

When considering sharing available key resources, most solutions rely on the application key rates as a metric. Each application receives only a proportional share of resources based on its requirements. As a result, in the case of limited resources and a large number of applications, the share that each application receives will be significantly lower than its requirements. In contrast, the SECOQC solution confirms the availability of resources before granting application access. However, in some cases, the provided key rate is adjusted to reflect the state of the network and the service is not guaranteed. The NSFC SDQaaS approach considers only the SKR of QKD links, not the number of available keys in storage. Each application is guaranteed the key rate it requires, allowing for the provision of a service of guaranteed quality, assuming that the performance of the QKD links is stable. However, not all applications, regardless of key rate requirements, should have equal priority in accessing resources. It is reasonable to assume that some applications are of greater critical importance and should thus receive higher priority. Priorities of application requirements were rarely considered in this context. The QoS-supported key manager distinguishes three service classes. Services are differentiated according to the time (delay) requirement. The highest priority class, the key-guaranteed service, is based on a resource reservation approach, whereas the other two classes use a queuing method. The Quantum Canada approach proposes classifying applications based on security requirements and identifies five categories. However, it appears that these classes are only used to identify fallback methods for each class rather than to prioritize requests. The NKPs DDKA-QKDN scheme takes a slightly different approach, prioritizing requests from the QKD network perspective. Request priority is determined by balancing security and key quantity requirements. Based on this analysis, it can be concluded that there is still no established approach to resource sharing and prioritizing requests to ensure varying levels of service for applications of different purposes. These functionalities are crucial, highlighting the current deficiencies in key manager capabilities, especially considering the growing trend of integrating QKD networks as enterprise services. 

The debate over whether fallback methods should be supported within QKD networks is ongoing. There's a growing interest in integrating PQC methods into key management systems. This approach allows for the utilization of both PQC and QKD methods for dual key agreement, providing a straightforward fallback option if QKD key resources become unavailable. However, it's crucial to assess the security requirements of the application and determine if they can be met with less secure cryptographic keys. The SwissQuantum solution adopts this approach, which is also listed in the ITU-T functional requirements for the key management layer. Quantum Canada's approach differs somewhat and entails significantly more complex key lifecycle management depending on the scenario. This raises the question: why should a QKD network concern itself with whether a key will be used to derive multiple session keys? Such discussions might be better suited for the service layer rather than the network layer.

\subsection{Secure key storage and key formatting}
\label{sec:discussion-storage}
The key manager is required to securely store and format keys where necessary for internal purposes or for key supply or key relay, including combining or splitting where lengths are not appropriate. The necessity for secure key storage has been acknowledged from the outset, owing to the unique characteristics of the QKD process. As a result, existing approaches studied in this article include this fundamental requirement of the key manager. Merely stating that keys will be stored is insufficient; the manner in which they are stored is equally crucial, as it greatly influences the effectiveness of key servicing capabilities. Table~\ref{tab:key-storage} provides a tabular comparison of existing key managers from the perspective of key storage realization and closely related techniques associated with this functionality.

\renewcommand{\arraystretch}{1}
\begin{table}
\centering
\caption{Comparison of the key managers in terms of key storage and related functionalities.}
\resizebox{0.98\textwidth}{!}{%
\begin{tabular}{lcccccccc}
\toprule
\multicolumn{1}{l}{Features}                                                 & \multirow{2}{*}{Year}                                   & \multirow{2}{*}{\begin{tabular}[c]{@{}c@{}}Layered\\ architecture\end{tabular}}                            & \multirow{2}{*}{\begin{tabular}[c]{@{}c@{}}Local key\\ storage\end{tabular}} & \multirow{2}{*}{\begin{tabular}[c]{@{}c@{}}Local keys\\ storage method\end{tabular}} & \multirow{2}{*}{Re-format}                                                     & \multirow{2}{*}{\begin{tabular}[c]{@{}c@{}}Logical division\\ of the keys\end{tabular}}                                                                           & \multirow{2}{*}{\begin{tabular}[c]{@{}c@{}}Global key\\ storage\end{tabular}} & \multirow{2}{*}{Storage thresholds}                                                                        \\ 
\multicolumn{1}{l}{Solutions}                                                                   &                                                         &                                                                                                            &                                                                              &                                                                                      &                                                                                &                                                                                                                                                                   &                                                                               &                                                                                                            \\ \midrule
\begin{tabular}[c]{@{}l@{}}DARPA key\\ management\end{tabular}                 & \begin{tabular}[c]{@{}c@{}}2002\\ -\\ 2007\end{tabular} & No                                                                                                         & Yes                                                                          & \begin{tabular}[c]{@{}c@{}}Fixed-sized\\ blocks\end{tabular}                         & No                                                                             & No                                                                                                                                                                & No                                                                            & No                                                                                                         \\ \hline
\begin{tabular}[c]{@{}l@{}}SECOQC key\\ management\end{tabular}                & \begin{tabular}[c]{@{}c@{}}2007\\ -\\ 2009\end{tabular} & \begin{tabular}[c]{@{}c@{}}Link\\ Network\\ Transport\\ Application\end{tabular}                           & Yes                                                                          & \begin{tabular}[c]{@{}c@{}}Homogeneous\\ block storage\end{tabular}                  & \begin{tabular}[c]{@{}c@{}}Yes - On reception;\\ No - On supply\end{tabular}   & \begin{tabular}[c]{@{}c@{}}Yes.\\ Keys for encryption\\ and decryption purposes\end{tabular}                                                                      & No                                                                            & \begin{tabular}[c]{@{}c@{}}Monitoring\\ of decryption\\ storage to trigger\\ refill procedure\end{tabular} \\ \hline
NIST manager                                                                   & 2008                                                    & No                                                                                                         & Yes                                                                          & \begin{tabular}[c]{@{}c@{}}Homogeneous\\ byte storage\end{tabular}                   & \begin{tabular}[c]{@{}c@{}}Yes. Supply in multiple\\ of bytes\end{tabular}     & \begin{tabular}[c]{@{}c@{}}Yes.\\ Session-based approach\end{tabular}                                                                                             & No                                                                            & No                                                                                                         \\ \hline
\begin{tabular}[c]{@{}l@{}}QCC security\\ processor\\ key manager\end{tabular} & 2008                                                    & No                                                                                                         & Yes                                                                          & -                                                                                    & -                                                                              & \begin{tabular}[c]{@{}c@{}}Yes.\\ Session-based approach\end{tabular}                                                                                             & No                                                                            & \begin{tabular}[c]{@{}c@{}}Threshold value\\ (fixed value) to start\\ refill procedure\end{tabular}        \\ \hline
\begin{tabular}[c]{@{}l@{}}NEC key\\ management\end{tabular}                   & 2009                                                    & \begin{tabular}[c]{@{}c@{}}Key generation\\ Connection\\ Key management\\ Communication\end{tabular}       & Yes                                                                          & \begin{tabular}[c]{@{}c@{}}Fixed-size\\ key files\end{tabular}                       & No                                                                             & \begin{tabular}[c]{@{}c@{}}Yes.\\ Keys for encryption\\ and decryption purposes\end{tabular}                                                                      & Yes                                                                           & No                                                                                                         \\ \hline
\begin{tabular}[c]{@{}l@{}}Magiq Tehnologies\\ key manager\end{tabular}        & 2010                                                    & \begin{tabular}[c]{@{}c@{}}QKD\\ Persistent storage\\ Key manager\\ Key storage\\ Application\end{tabular} & Yes                                                                          & \begin{tabular}[c]{@{}c@{}}Fixed-sized\\ blocks\end{tabular}                         & No                                                                             & \begin{tabular}[c]{@{}c@{}}Yes.\\ Session-based approach\end{tabular}                                                                                             & No                                                                            & No                                                                                                         \\ \hline
\begin{tabular}[c]{@{}l@{}}SwissQuantum\\ key management\end{tabular}          & \begin{tabular}[c]{@{}c@{}}2009\\ -\\ 2011\end{tabular} & \begin{tabular}[c]{@{}c@{}}Quantum\\ Key management\\ Application\end{tabular}                             & Yes                                                                          & -                                                                                    & -                                                                              & \begin{tabular}[c]{@{}c@{}}Yes.\\ Session-based approach\end{tabular}                                                                                             & Yes                                                                           & -                                                                                                          \\ \hline
\begin{tabular}[c]{@{}l@{}}QoS-supported\\ key manager\end{tabular}            & 2011                                                    & No                                                                                                         & Yes                                                                          & -                                                                                    & -                                                                              & -                                                                                                                                                                 & -                                                                             & -                                                                                                          \\ \hline
\begin{tabular}[c]{@{}l@{}}NECTEC\\ key management\end{tabular}                & 2012                                                    & \begin{tabular}[c]{@{}c@{}}Quantum\\ Key management\\ Application\end{tabular}                             & Yes                                                                          & -                                                                                    & -                                                                              & \begin{tabular}[c]{@{}c@{}}Yes.\\ Session-based approach\end{tabular}                                                                                             & Yes                                                                           & No                                                                                                         \\ \hline
\begin{tabular}[c]{@{}l@{}}Toshiba\\ key management\end{tabular}               & 2016                                                    & \begin{tabular}[c]{@{}c@{}}Link\\ Network\\ Transport\\ Application\end{tabular}                           & Yes                                                                          & \begin{tabular}[c]{@{}c@{}}Homogeneous\\ block storage\end{tabular}                  & \begin{tabular}[c]{@{}c@{}}Yes - On reception;\\ No - On supply\end{tabular}   & \begin{tabular}[c]{@{}c@{}}Yes.\\ Keys for encryption\\ and decryption purposes \\ combined with a \\ session-based approach\end{tabular}                         & Yes                                                                           & No                                                                                                         \\ \hline
\begin{tabular}[c]{@{}l@{}}NICT QKD\\ platform\end{tabular}                    & 2017                                                    & \begin{tabular}[c]{@{}c@{}}Quantum\\ Key management\\ Key supply\\ Application\end{tabular}                & Yes                                                                          & -                                                                                    & -                                                                              & -                                                                                                                                                                 & Yes                                                                           & No                                                                                                         \\ \hline
\begin{tabular}[c]{@{}l@{}}Quantum Canada\\ key management\end{tabular}        & 2018                                                    & \begin{tabular}[c]{@{}c@{}}Link\\ Network\\ Key management service\\ Host\end{tabular}                     & Yes                                                                          & -                                                                                    & -                                                                              & \begin{tabular}[c]{@{}c@{}}Yes.\\ Keys for local supply \\ and keys for global key\\ distribution.\\ Keys for encryption\\ and decryption\\ purposes\end{tabular} & Yes                                                                           & No                                                                                                         \\ \hline
\begin{tabular}[c]{@{}l@{}}NSFC SDQaaS\\ framework\end{tabular}                & 2019                                                    & \begin{tabular}[c]{@{}c@{}}Infrastructure\\ Control\\ Application\end{tabular}                             & Yes                                                                          & -                                                                                    & -                                                                              & \begin{tabular}[c]{@{}c@{}}Yes.\\ Session-based approach\end{tabular}                                                                                             & Yes                                                                           & No                                                                                                         \\ \hline
\begin{tabular}[c]{@{}l@{}}NKPs \\ DDKA-QKDN\\ scheme\end{tabular}             & 2022                                                    & No                                                                                                         & Yes                                                                          & \begin{tabular}[c]{@{}c@{}}Five predetermined\\ fixed-sized blocks\end{tabular}      & \begin{tabular}[c]{@{}c@{}}Yes - On reception;\\ No - On supply\end{tabular}   & No                                                                                                                                                                & Yes                                                                           & \begin{tabular}[c]{@{}c@{}}Threshold value to\\ start global key\\ distribution\end{tabular}               \\ \hline
\begin{tabular}[c]{@{}l@{}}KISTI\\ key management\end{tabular}                 & 2022                                                    & \begin{tabular}[c]{@{}c@{}}Quantum\\ Key management\\ Key supply\\ Application\end{tabular}                & Yes                                                                          & -                                                                                    & Yes - Using HKDF                                                               & -                                                                                                                                                                 & Yes                                                                           & No                                                                                                         \\ \hline
\begin{tabular}[c]{@{}l@{}}AIT\\ key manager\end{tabular}                      & 2023                                                    & No                                                                                                         & Yes                                                                          & -                                                                                    & \begin{tabular}[c]{@{}c@{}}Yes. Split and merge\\ methods defined\end{tabular} & -                                                                                                                                                                 & -                                                                             & No                                                                                                         \\ \hline
\begin{tabular}[c]{@{}l@{}}Cisco\\ key manager\end{tabular}                     & 2023                                                   & No                                                                                                         & Yes                                                                          & \begin{tabular}[c]{@{}c@{}}Fixed-sized\\ blocks (64 bits)\end{tabular}                        & Yes                                                                                                               & No                                                                           & -                                                                            & \begin{tabular}[c]{@{}c@{}}Threshold value\\ (fixed value) to start\\ refill procedure\end{tabular}                                                       \\ \bottomrule
\end{tabular}
}
\label{tab:key-storage}
\end{table}
\renewcommand{\arraystretch}{1}

An in-depth analysis of existing approaches has uncovered several prevailing key storage designs, alongside numerous shortcomings that require attention and resolution. To address the inefficiencies in key delivery identified within the DARPA quantum network, the SECOQC approach (Section~\ref{sec:sqcoqc}) suggests categorizing keys based on their intended purpose, separating them into encryption and decryption keys. This approach allows key managers to facilitate the seamless utilization of keys for delivery or key relaying, eliminating concerns about disagreements or collisions in key access. Cryptographic applications, as well as internal processes like key relaying, have shared access to the singular encryption key storage. The decryption key storage, however, is accessed only upon instruction from the peer key manager or when a slave cryptographic application requests the service. Given the significance of the SECOQC architecture, this key management approach is adopted in Toshiba, and it's highly probable that it's also utilized in the NICT QKD platform. This assumption is based on the fact that the NICT QKD platform is built upon the work developed for the Tokyo QKD network, which employed the same architecture as SECOQC. Similarly, the following solutions define singular encryption and decryption key storages: NEC (Section~\ref{sec:nec}), and Quantum Canada (Section~\ref{sec:canada}). The encryption and decryption storages in SECOQC (and thus, Toshiba) contain predefined-size key blocks that are sequentially stored. This approach makes it straightforward to reformat keys that arrive from QKD modules in large and varying sizes into blocks of predefined sizes. However, it also implies that keys are suitable for use/supply in an ordered manner and in a single predefined size -- the block size. The block size, however, has not been discussed. Similarly, NEC's key management solution defines keys as fixed-size files.

NIST (Section~\ref{sec:nist}) introduced a session-based key storage approach, which is used in several solutions, including the QCC security processor (Section~\ref{sec:qcc}), Magiq Technologies (Section~\ref{sec:magiq}), SwissQuantum (Section~\ref{sec:swiss}), NECTEC (Section~\ref{sec:nectec}), and NSFC SDQaaS framework (Section~\ref{sec:sdqaas}). Each cryptographic application is assigned its own key storage and a subset of the available keys from the common storage. Each application typically registers with a desired key rate, which serves as a guideline for assigning available keys to multiple buffers. One of the packet scheduling algorithms can be used to ensure that keys are assigned fairly across various buffers. 
The NIST defines session-based storage as containing keys that are reformatted to a one-byte size and stored sequentially. This method is appropriate for meeting the varying key size requirements of cryptographic applications, as any size in bits (that is a multiple of 8) can be supplied by combining multiple bytes from storage. The Magiq Technologies key management approach specifies that keys are stored in predetermined fixed-size blocks, but the block size is unknown. Because the reformatting is not specified, the keys are most likely provided in block format.

Although the discussed approaches to key storage are primarily for local keys, the same constructions could be used for global keys as well. However, it is worth noting that the approach to global key storage is rarely discussed. The SECOQC makes it abundantly clear that global key storage and management are not supported. Upon completion of global key distribution, keys are simply provided to the requesting application. Given that global keys are distributed in larger blocks to reduce encryption overhead, the system may be inefficient because large keys are supplied to the application regardless of its size requirements. This transfers responsibility for global key management to the application that will use the obtained key. The Toshiba defines the global key management function but does not discuss the storage method.

The deficiency of key managers becomes apparent in solutions that store keys as fixed-sized blocks, whether they are in shared encryption storages or dedicated application storages. This oversight occurs solely because of neglecting key formatting when requested sizes do not align with block sizes. For example, the key management systems of DARPA and Magiq Technologies clearly states that the keys are received from the quantum layer at a predetermined size and thus stored without any reformatting. Since the key formatting is not supported at supply, the system is expected to have low efficiency as large chunks of keys are delivered on any request, even requests that require tiny quantities. NKP DDKA-QKDN scheme partially addresses this problem by storing keys in five different but predetermined sizes, namely 128, 256, 512, 1024, and 2048 bits. This means the reformat function is executed before storage. However, it does not define the ratio of how many keys and of which size should be created at a given time. More sophisticated ways of solving this problem have been proposed by AIT and partly by KISTI by introducing the function of reformatting on supply. By having this function, it can be concluded that both are indeed store keys in fixed-sized blocks. KISTI uses one approach considered controversial in terms of QKD networks, namely the use of the HKDF function to derive smaller keys, which are supplied on demand, from large keys. Because many smaller keys are derived from a single key, they are not eligible for the ITS profile. The AIT defines split and merge methods for reformatting available keys to desired sizes at the key management layer. This solution can apply to all solutions that lack this functionality. The Cisco key manager stores keys in fixed-size 64-bit blocks, supporting the reformat function before storage. However, it does not address how to create the supply key of the requested size. The key may indeed be of any length—in this case, it may be a multiple of 64—for the application mentioned in section~\ref{sec:cisco}, where it is used in the key derivation procedure.

Some of the mentioned research works discuss storage thresholds, which are important for improving efficiency and enabling continuous supply without interruption. The SECOQC key management approach monitors the available number of decryption keys and requires a refill procedure if this number is critically low. Since the keys used for decryption have encryption copies on another node, timely refilling allows for continuous transfer of sensitive data. Similarly, the QCC key manager applies thresholds to a session-based approach. Suppose the number of keys in the application's dedicated buffer falls below a threshold value. In that case, new keys, if available, are assigned promptly so that the application does not experience interruptions in key supply.
Similarly, the Cisco key manager defines a threshold value that causes new keys to be pulled from the QKD devices. Threshold values can be assigned to global key storage, as in NKP's DDKA-QKDN scheme. This allows the distribution of a sufficient number of global keys in advance. Global key storage thresholds are more important and complex than previously discussed thresholds introduced in SECOQC and QCC solutions. Since the global key distribution time can vary, the lower threshold should be set dynamically to account for this distribution time.

From this discussion, it's apparent that the methods of storing and managing keys within the key management system are still notably constrained. Furthermore, while we can identify various approaches to key storage, there is a lack of research examining the comparative effectiveness of these approaches. A system that supports both identified key storage approaches—encryption and decryption, and session-based—would be capable of accommodating both ETSI interfaces for accessing services. A shared encryption key storage would cater to ETSI 014 requests, while the session-based approach would respond to the session-based nature of the ETSI 004 interface. Additionally, many solutions lack a critical key formatting capability, which may compromise the QKD service. Applications often receive key blocks larger in size than required, reducing the probability of serving other applications due to decreased key material in storages. The initial design of the AIT key manager indicates support for merge and split operations. However, questions persist regarding the efficiency with which these operations can be executed. Moreover, there is a significant gap in supporting a large number of applications with guaranteed service levels and ensuring fair sharing of scarce resources. Presently, if supported at all, solutions offer a ratio of available key material based on desired application rates. There is no assurance of a guaranteed level of service. For technology to be suitable for application in critical infrastructures, it's crucial to have a method of differentiating application priorities and ensuring guaranteed levels of service.

\section{Key challenges and future directions}
\label{sec:challenges}
This section presents a concise list of challenges and future endeavors in key management in QKD networks. Key management is the foundational task of QKD networks; without it, the key generation process would be ineffective. However, the development and optimization of the key management layer are significantly constrained due to the limited number of testbeds and real-world applications of QKD network services. These are identified key challenges to guide future directions:

\begin{itemize}
    \item \textbf{Standardization of interfaces for communication between QKD modules and key managers:} To facilitate fair representation of all QKD equipment manufacturers in the market, it's imperative to establish clear guidelines for communication and interoperability between various QKD module manufacturers and key managers. Currently, there is no standardized interface that delineates this communication.

    \item \textbf{Effective key storage approaches:} The key manager is required to effectively manage cryptographic keys, which includes key storage and formatting. It is critical that keys are stored in a manner that allows for efficient key addition and retrieval. Key formatting should also be supported and performed efficiently on demand because the application in critical infrastructures requires a minimum delay in key supply. There are no studies analyzing the effectiveness of key storage designs, nor are there detailed approaches to effective key formatting.

    \item \textbf{Effective resource sharing and quality of service support:} With the growing push for enterprise integration of QKD networks, it's anticipated that a large number of users, i.e., cryptographic applications, will need to share limited key resources. The key manager must prioritize applications of varying backgrounds and even support quality of service requirements for the most critical applications. Most approaches utilize desired key rates as a metric for allocating available resources. However, it's essential to consider the nature of the applications.

    \item \textbf{Standardization of interfaces for communication between remote key managers:} To achieve interoperability and ensure fair representation of different key manager vendors, it's crucial to define the methods and protocols between remote key managers. This point presents a significant obstacle to developing and integrating QKD networks. As vendors endeavor to incorporate their QKD devices and key managers into a single device and rely on proprietary interfaces, the result is the imposition of a QKD network reliant on the catalog of a single vendor. The absence of a standardized interface between key managers hinders the broader integration of QKD networks, particularly in terms of key relaying, which is aimed to be addressed through the ETSI QKD 020 standard. However, there are concerns that this solution may not scale effectively and could potentially slow down the traffic within the QKD network.
    
\end{itemize}

\section{Conclusion}
\label{sec:conclusion}

The key management layer must be addressed for QKD networks to become a viable technology. It enables the realization of QKD networks by overcoming the point-to-point limitations of QKD links. It determines the QKD network's ability to provide a service with guaranteed Quality of Service (QoS) and delivery of keys to end users safely and on time. Finally, it enables the interoperability of QKD equipment by connecting different types of QKD links into a single QKD network.
This paper extensively reviews the evolution of key managers in QKD networks. It analyzes and compares existing solutions regarding key storage and service provisions. To the best of our knowledge, this is the first paper to examine approaches to developing the key manager component in QKD networks. The main contribution of this paper is an in-depth analysis of existing approaches for developing key management systems as fundamental components of the QKD network.

\section*{Acknowledgments}
    The research leading to the published results was supported by the Ministry of the  Interior of the  Czech  Republic under grant ID VJ01010008  within the project  Network  Cybersecurity in Post-Quantum  Era, partly by the NATO SPS G5894 project "Quantum Cybersecurity in 5G Networks (QUANTUM5)". This work was also supported by the Ministry of Science, Higher Education and Youth of Canton Sarajevo, Bosnia and Herzegovina under Grant No. 27-02-35-37082-1/23, within the project DQKDNM 2023.

\bibliographystyle{unsrtnat}


\end{document}